\documentclass[aps,prd,twocolumn,superscriptaddress,showpacs]{revtex4-1}

\usepackage{amssymb}
\usepackage{amsmath}
\usepackage{graphicx}
\usepackage{enumerate}
\usepackage{mathtools}
\begin{document}

%\preprint{}

\title{Geometric phases in neutrino oscillations with nonlinear refraction}


\author{Lucas Johns}
%\thanks{}
\author{George M. Fuller}
\affiliation{Department of Physics, University of California, San Diego, La Jolla, California 92093, USA}


\date{\today}

\begin{abstract}
Neutrinos propagating in dense astrophysical environments sustain nonlinear refractive effects due to neutrino--neutrino forward scattering.  We study geometric phases in neutrino oscillations that arise out of cyclic evolution of the potential generated by these forward-scattering processes.  We perform several calculations, exact and perturbative, that illustrate the robustness of such phases, and of geometric effects more broadly, in the flavor evolution of neutrinos.  The scenarios we consider are highly idealized in order to make them analytically tractable, but they suggest the possible presence of complicated geometric effects in realistic astrophysical settings.  We also point out that in the limit of extremely high neutrino densities, the nonlinear potential in three flavors naturally gives rise to non-Abelian geometric phases.  This paper is intended to be accessible to neutrino experts and non-specialists alike.
\end{abstract}

%\keywords{}

\maketitle

%\section{}
%\subsection{}
%\subsubsection{}

\section{Introduction \label{introsec}}

Geometric phases in neutrino propagation have been investigated in various guises \cite{nakagawa1987, vidal1990, aneziris1991, smirnov1991, akhmedov1991, guzzo1992, naumov1992, naumov1994, blasone1999, wang2001, he2005, blasone2009, mehta2009, joshi2016} over the decades since it was recognized that neutrino flavor transformation might provide the solution to the mysterious deficit of solar neutrinos \cite{mikheyev1985, bethe1986}.  In the intervening years the understanding of how neutrino oscillations are modified in medium has undergone a sea change.  In particular, it is now recognized that in environments with very high neutrino density the flavor evolution of one neutrino is coupled to that of all other neutrinos with which it interacts.  The result is a colorful tapestry of flavor-transformation phenomena that extends far beyond the classic resonance mechanism at work in solar neutrinos.

In this paper we conduct the first study of neutrino geometric phases that accounts for the nonlinear coupling of flavor states, a phenomenon known in the neutrino literature as \textit{self-coupling}. The phases that we exhibit in our calculations persist at the probability level and are therefore detectable in principle, though no attempt is made here to extract geometric phases from models of any degree of astrophysical realism.  By and large such models would necessitate numerical analysis, which may obscure some of the insights otherwise made transparent by an analytical treatment.  Our present aim is to explore the manifestations of geometric phases precisely without the complications that continue to make the modeling of neutrino flavor such a disobliging task.  Even so, as we argue here, one gleans a hint that geometric quantum effects of one form or another may be nearly unavoidable in the flavor evolution of neutrinos in such dense environments as core-collapse supernovae or neutron-star mergers.

Broadly, geometric phase refers to the extra, path-dependent quantum phase that a state accumulates in addition to the dynamical phase from the ``local'' influence of the Hamiltonian. The latter phase is present even for a time-independent Hamiltonian and its importance has been appreciated since the very advent of quantum mechanics; we denote it by $\delta$, and for a state $| \psi \rangle$ and Hamiltonian $H$ it has the usual form
\begin{equation}
\delta = - \int dt ~ \langle \psi (t) | H ( t ) | \psi (t) \rangle. \label{dynamicaldef}
\end{equation}
The appreciation of geometric phases as a common and observable feature of many quantum systems is much more recent, dating back to the seminal realization by Berry \cite{berry1984} that a state acted on by a cyclic, adiabatically changing Hamiltonian acquires a phase whose value depends on the circuit traced out by the Hamiltonian in the space of its parameters.  If $| \psi \rangle$ begins as an instantaneous energy eigenstate $| \eta \rangle$, the adiabatic theorem dictates that it remains so, \textit{i.e.}, $| \psi (t) \rangle = e^{i \phi (t)} | \eta (t) \rangle$, and the total phase has the form $\phi = \delta + \gamma$ with the geometric phase given by $\gamma$.  After the system, which is described by time-dependent parameters $\vec{R} (t)$, completes a circuit $\mathcal{C}$, the state $| \psi \rangle$ has developed a geometric phase
\begin{equation}
\gamma = i \oint _\mathcal{C} \langle \eta (t) | \nabla | \eta (t) \rangle \cdot d \vec{R}, \label{gammadef}
\end{equation}
where $\nabla$ is the gradient operator in $\vec{R}$-space.

The particular incarnation of the geometric phase defined by this expression is often called the \textit{Berry phase} and is specific to cyclic, adiabatic systems.  The notion can be generalized enormously: to entangled states \cite{sjoqvist2000a}; to mixed states \cite{sjoqvist2000b}; and to non-adiabatic \cite{nakagawa1987, aharonov1987, anandan1988}, non-cyclic \cite{samuel1988, pati1998, polavieja1998a, polavieja1998b, mostafazadeh1999}, and even open or non-Hamiltonian \cite{garrison1988, gamliel1989, dattoli1990, sun1993, carollo2003} systems.  In this paper we use the broad term \textit{geometric phase} but our typical targets are indeed Berry phases of the form in Eq.~\eqref{gammadef}.  At several points we will make contact with some of these generalizations.

On an intuitive level the existence of geometric phases in quantum systems is perhaps most immediately grasped by analogy to the classical world: A Foucault pendulum, carried around a closed loop on the surface of the Earth in such a way that the plane of oscillation is never rotated, nonetheless returns to its starting point with a rotated plane of oscillation, and the angle by which the plane has rotated (known as the Hannay angle) is, moreover, equal to the solid angle enclosed by the loop.  Quantum geometric phases are rotations of the wavefunction and arise, in much the same way, from parallel transport along a path.  For readers seeking a deeper understanding of the phenomenon, we sketch a picture in Sec.~\ref{geombackground} of geometric phases from the perspective of differential geometry, which permits a rigorous translation of this classical intuition into the quantum realm.  The picture is also intimately related to the ``polarization vectors'' (formally introduced in Sec.~\ref{neubackground}) that are ubiquitous in the literature on neutrino flavor evolution.  Put succinctly, this connection is why geometry is relevant to neutrino oscillations.

The observability of geometric phases is today well-established in a variety of settings thanks to such landmark experiments as those in Refs.~\cite{tomita1986, bitter1987, suter1987, tycko1987, bhandari1988, bird1988, chyba1988, suter1988, weinfurter1990} and the vast sweep of investigations carried out in more recent years.  The modern understanding of geometric phases has also shed light on instances and variations of the phenomenon that were predicted or observed \textit{before} Berry's original analysis, perhaps the most famous cases being the Aharonov--Bohm effect \cite{aharonov1959}, the Pancharatnam phase \cite{pancharatnam1956}, and the (retroactively named) molecular Berry phase \cite{herzberg1963, mead1979}.  

Interferometry experiments, which comprise a large share of the corpus of geometric-phase studies, are plainly out of the question when it comes to neutrinos.  But neutrino oscillations are themselves fundamentally an interference phenomenon: As mass eigenstates propagate in vacuum they pick up phases at different rates, and the interference between these phases gives rise to flavor oscillation.  It is natural, then, to wonder whether the interference intrinsic to neutrino oscillations might function in some way as an ``interferometer'' sensitive to geometric phases.

Much of the earliest interest in this possibility surrounded the idea that the resolution of the missing-solar-neutrinos puzzle may come from the conversion of neutrinos into antineutrinos via the interaction of their magnetic moments with solar magnetic fields \cite{cisneros1971, okun1986, vidal1990, smirnov1991}.  Although the consensus is that the Mikheyev--Smirnov--Wolfenstein (MSW) mechanism \cite{wolfenstein1978, mikheyev1985} ultimately won the day as far as the solar neutrino problem, neutrino geometric phases continue to be explored in the context of astrophysical magnetic fields \cite{joshi2016}.  Furthermore, just as magnetic field vectors can trace out closed loops in physical space, the optical potentials generated by coherent forward scattering of neutrinos with background particles raise the possibility that a similar mechanism might operate in flavor space.

%A key idea that we advocate in this paper is that neutrinos pose an interesting setting for geometric phases because multiple bases are important for describing their behavior in medium.  The upshot is that even if geometric phases only appear in energy eigenstates at the amplitude level, traces of their presence may be visible in the transition probabilities between \textit{flavor} eigenstates.  This observation may be of importance in situations where neutrinos are produced as (or decouple into) flavor eigenstates and subsequently undergo coherent adiabatic evolution under the influence of a cyclic Hamiltonian. \textbf{[do something with this paragraph]}

Shortly after the explosion of interest in geometric phases began, Nakagawa \cite{nakagawa1987} acknowledged this possibility but observed that geometric phases cannot appear in two-flavor neutrino oscillations in a matter background, for the simple reason that there is just one parameter in the Hamiltonian that is varying (viz., the density of matter particles) and therefore a cycle of finite area cannot be traced out.  Naumov \cite{naumov1992, naumov1994} later showed that geometric phases can emerge in three flavors, provided that there is both CP violation and a cyclically varying number density of scatterers.  These papers regarded vacuum mixing as the sole contributor to the off-diagonal (in the flavor basis) Hamiltonian matrix elements; Pantaleone's insight \cite{pantaleone1992} that self-coupling can also supply off-diagonal contributions was not yet widely appreciated.

More recently geometric phases were studied by He \textit{et al.} \cite{he2005} in a paper generalizing Naumov's work to active--sterile mixing and nonstandard interactions.  Although the authors noted that the neutrino-background density is an additional parameter varying independently of the matter-background density, they did not consider the contribution of coherent neutrino--neutrino scattering to the off-diagonal Hamiltonian elements.  We thus point out for the first time in the literature that geometric phases can arise out of the self-coupling potential and can appear even with just two flavors.  We also argue that because of the nonlinear nature of this potential, neutrino self-coupling in flavor space is a particularly rich avatar for geometric phases.

%Pursuing a different angle, the papers by Blasone \textit{et al.} \cite{blasone1999} and by Wang \textit{et al.} \cite{wang2001} examined geometric phases in vacuum oscillations; but in these contexts the geometric parameters are the vacuum mixing angles and there are no cyclic external parameters, and what they mean by ``geometric phase'' is very different from what we mean.  \textbf{[Aharanov--Anandan phase?]}  More recently, Mehta \cite{mehta2009} extended this line of thinking. \textbf{[rewrite this paragraph]}

Our approach is to perform calculations on several toy models that reveal various facets of geometric phases in the presence of nonlinear neutrino--neutrino coupling.  The calculations that follow shed light on the precise role of adiabaticity, the nonlinear entangling of the geometric phases developed by neutrinos in interaction with one another, the fragile cyclicity of flavor transformation, and the non-Abelian phase structure of a certain three-flavor limit.  While we do not attempt to locate geometric phases in realistic astrophysical models, our results are suggestive of the prevalence in sophisticated numerical computations of geometric effects generally, if not specifically the cyclic, adiabatic phases we investigate.

In Sec.~\ref{geombackground} we sketch the picture of geometric phases from the viewpoint of differential geometry.  In Sec.~\ref{neubackground} we present the relevant background on medium-enhanced neutrino oscillations.  We then turn our attention to the flavor evolution that occurs in a system of two coupled neutrino populations.  Working in the two-flavor approximation, we examine three limiting cases: the mixed-potential limit in Sec.~\ref{2mixedsec}, the pure-self-coupling limit in Sec.~\ref{puresec}, and the weak-self-coupling limit in Sec.~\ref{weakselfsec}.  After examining these two-flavor scenarios, we return in Sec.~\ref{3mixedsec} to the mixed-potential limit, this time in three flavors, and show how it begets non-Abelian geometric phases.  We conclude in Sec.~\ref{concsec}.

\section{The differential-geometric picture \label{geombackground}}

The Born rule implies that the overall phase of a quantum state $| \psi \rangle$ is inessential for computing or measuring observables at some time $t$.  With a rephasing $| \tilde{\psi} (t) \rangle = e^{i\alpha} | \psi (t) \rangle$ by some arbitrary phase $\alpha$,
\begin{equation}
\langle \tilde{\psi} (t) | \mathcal{O} | \tilde{\psi} (t) \rangle = \langle \psi (t) | \mathcal{O} | \psi (t) \rangle
\end{equation}
for any Hermitian operator $\mathcal{O}$.  As a result the value of $\alpha$ in $| \tilde{\psi} (t) \rangle$ can be chosen arbitrarily, but as shown by Berry it does not follow that the overall phase can be ignored altogether.  At a basic level, geometric phases are in fact amenable to observation because once $\alpha$ is chosen for $| \tilde{\psi} (t) \rangle$, the phase of $| \tilde{\psi} (t') \rangle$ at any other time $t'$ is predetermined by this choice (in conjunction, of course, with the dynamics of the system).  In other words, phase \textit{changes} are physically significant.

Granting that geometric phases exist, it is perhaps not so surprising that they are observable---but that they should exist at all is a profound fact about quantum mechanics.  Fundamentally the existence of geometric phases is a consequence of Hilbert space having nontrivial geometry, and the values of the phases are governed by that geometry in tandem with the relevant Hamiltonian.  Just as the spheroidal shape of the Earth determines the Hannay angle \cite{hannay1985} of a Foucault pendulum carried along the planet's surface, in an analogous way does the ``shape'' of Hilbert space determine how a wavefunction rotates---that is, picks up phase---as it is moved along a path.

These ideas are most naturally expressed in a rigorous manner using the language of principal fiber bundles and their associated structures.  The relation between geometric phases and fiber bundles was the powerful insight of Simon \cite{simon1983} and has been elaborated by many subsequent authors.  We now try to elucidate this helpful way of understanding geometric phases.

For a Hamiltonian $H [\vec{R}]$ that depends on the time-dependent parameters $\vec{R} (t)$, a given nondegenerate instantaneous eigenstate $| \tilde{\psi} \rangle$ can be specified by the pair $\left( \vec{R}, \exp (i\alpha) \right)$, where $\alpha$ is the arbitrary overall phase referred to previously.  We have already established that this pair corresponds to the same physical state regardless of the value of $\alpha$; in technical terms, the pair projects down to the same ray in projective Hilbert space for all $\alpha$.  Hilbert space can thus be visualized as a manifold with an identical string piercing through every point: A given point on the manifold corresponds to a physical state (or, equivalently, a value of $\vec{R}$) and the string through that point represents the possible choices of phase $\alpha$ for that particular physical state.

\begin{figure}
\includegraphics[width=.48\textwidth]{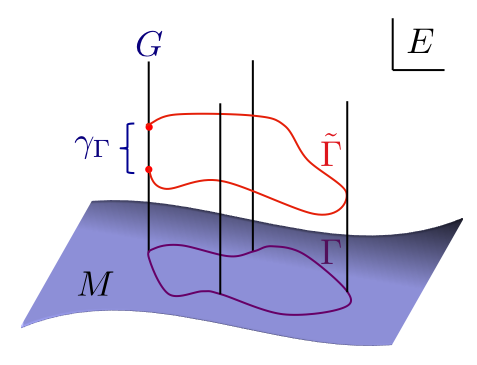}
\caption{A visualization of the differential-geometric structures underpinning the geometric phase.  The curve $\tilde{\Gamma}$ in Hilbert space projects down to a curve $\Gamma$ in projective Hilbert space.  $\tilde{\Gamma}$ begins and ends on the same fiber (labeled $G$, denoting the Lie group) but may not return to the same element in the fiber.  The group element at the start of the loop is brought to the element at the end by the holonomy $\gamma_\Gamma$.  The holonomy is precisely the geometric phase acquired upon completing the cycle.  \label{pfb}}
\end{figure}

A principal fiber bundle describes precisely such an object.  It consists of three parts (Fig.~\ref{pfb}): a total space $E$, a base manifold $M$, and a Lie-group fiber $G$.  The translation into bundle language uses the following associations:
\begin{align}
& E ~~~ \longleftrightarrow ~~ \textrm{States in Hilbert space,} \notag \\
& \hspace{0.8 in} \textrm{identified by $( \vec{R}, \exp (i\alpha) )$}, \notag \\
& M ~~ \longleftrightarrow ~~ \textrm{Physical states in projective} \notag \\
& \hspace{0.8 in} \textrm{Hilbert space}, \notag \\
& G ~~~ \longleftrightarrow ~~ \textrm{Elements $e^{i\alpha}$ of the group $U(1)$.}
\end{align}
If the eigenstate is part of an $n$-degenerate subspace, then the fiber $G$ is instead the non-Abelian group $U(n)$.  For simplicity we will continue to assume non-degeneracy throughout the remainder of this section, but we will demonstrate the emergence of non-Abelian fibers in the context of neutrino oscillations in Sec.~\ref{3mixedsec}.

As $\vec{R} (t)$ evolves, a curve $\mathcal{C}(t)$ in parameter space, hence also a curve on the base manifold $M$, is traced out.  The curve on $M$, $\Gamma (t)$, can be visualized as the shadow of a curve $\tilde{\Gamma} (t)$ on $E$: At every point in $M$ there is a point in $E$ elevated above the base manifold according to its fiber element.  The freedom to choose the value of $\alpha$ for $| \tilde{\psi} (t_i) \rangle$ at a specific time $t_i$ is manifested as a freedom to choose the fiber element at $\tilde{\Gamma} (t_i)$, but having made this choice, the fiber elements are non-arbitrary for $t > t_i$ (and, for that matter, for $t < t_i$).  For us, the condition of adiabaticity furnishes the principal fiber bundle with a connection, which is to say a way of moving a state from one fiber to the next.  The adiabatic connection is equivalent to the condition
\begin{equation}
\bigg\langle \tilde{\psi} (t) \bigg| \frac{d}{dt} \bigg| \tilde{\psi} (t) \bigg\rangle = 0. \label{parallel}
\end{equation}
This constraint describes the parallel transport of the state along the path and captures the intuitive notion that the state vector moves in such a way that locally it never appears to be rotating, modulo the dynamical phase evolution (Eq.~\eqref{dynamicaldef}).

If $\vec{R} (t_f) = \vec{R} (t_i)$, then $\Gamma$ forms a closed loop and the state returns at $t_f$ to the same fiber it began on at $t_i$.  It is not guaranteed, however, that the state will return to the same fiber \textit{element}.  The difference
\begin{equation}
e^{i \left[ \alpha (t_f) - \alpha (t_i) \right]} \equiv e^{i \Delta \alpha}
\end{equation}
is itself an element of the Lie group and is termed the holonomy of the connection on the principal fiber bundle.  It is precisely equal to the geometric phase: $\Delta \alpha = \gamma$.  The plausibility of a global rotation occurring without any local rotation can be seen from the non-transitivity of phase \cite{pancharatnam1956, anandan1992}: $\langle \tilde{\psi}_1 | \tilde{\psi}_2 \rangle$ and $\langle \tilde{\psi}_2 | \tilde{\psi}_3 \rangle$ having the same phase does not necessarily imply that $\langle \tilde{\psi}_1 | \tilde{\psi}_3 \rangle$ has the same phase as well.

The numerical value of the holonomy depends on the curve $\Gamma$ and the geometry of the manifold $M$ on which $\Gamma$ lies.  To illustrate this point concretely, we consider the evolution of two-flavor neutrinos in flavor space.  As with any two-level system, wavefunction normalization and the arbitrariness of $\alpha (t_i)$ relegate a $\mathbb{C}^2$ state vector to the Bloch sphere $S^2$.  In two flavors, therefore, it is the geometry of the Bloch sphere that determines the geometric phase associated with passage along a closed loop $\mathcal{C}$ in parameter space.  The principal fiber bundle can be pictured in this case as a ball ($M = S^2$) with spikes ($G = U(1)$) sticking out of it.  The three-flavor case with two degenerate eigenstates, which we turn to toward the end of this study, has base manifold $\mathbb{C}P^2 = SU(3) / U(2)$ and is not as easily visualized, but the message is the same: The geometry of Hilbert space leaves its footprint in the flavor conversion of neutrinos.

%As with any two-level system, wavefunction normalization and the arbitrariness of $\alpha (t_i)$ relegate a $\mathbb{C}^2$ state vector to the complex projective line $\mathbb{C}P^1$, represented through Hopf fibration by the Bloch (or Riemann, or Poincar\'e) sphere $S^2$.

\section{Neutrino oscillations in medium \label{neubackground}}

Neutrino oscillations are conveniently studied by tracking the time-evolution of the flavor wavefunction
\begin{equation}
| \psi \rangle = \left( \begin{array}{c}
a_{e} \\
a_{\mu} \\
a_{\tau}
\end{array} \right), \label{wfdef}
\end{equation}
where $a_{\alpha}$ is the amplitude for the neutrino to have $\alpha$ flavor.  This same state, expressed above in the flavor basis, can be translated into the mass basis via the Pontecorvo--Maki--Nakagawa--Sakata (PMNS) matrix $U_{\textrm{PMNS}}$:
\begin{equation}
| \psi \rangle _\textrm{f} = U_\textrm{PMNS} | \psi \rangle _\textrm{m},
\end{equation}
where the subscript denotes the basis.  (The wavefunction in Eq.~\eqref{wfdef} is really $| \psi \rangle _\textrm{f}$, but we will be dropping the subscripts as we proceed, leaving that job to the context.)  The mixing matrix is traditionally parameterized as
\begin{widetext}
\begin{equation}
U_\textrm{PMNS} = \left( \begin{array}{ccc}
c_{12} c_{13} & s_{12} c_{13} & s_{13} e^{-i \delta} \\
- s_{12} c_{23} - c_{12} s_{23} s_{13} e^{i \delta} & c_{12} c_{23} - s_{12} s_{23} s_{13} e^{i \delta} & s_{23} c_{13} \\
s_{12} s_{23} - c_{12} c_{23} s_{13} e^{- i\delta} & - c_{12} s_{23} - s_{12} c_{23} s_{13} e^{i \delta} & c_{23} c_{13}
\end{array} \right),
\end{equation}
\end{widetext}
using the notation $c_{ij} \equiv \cos \theta_{ij}$ and $s_{ij} \equiv \sin\theta_{ij}$ in terms of the oscillation angles $\theta_{12}$, $\theta_{23}$, and $\theta_{13}$.  The parameter $\delta$ is the Dirac CP-violating phase.  Measurements of the three mixing angles represent major triumphs of experimental particle physics over the past two decades; the Dirac phase, meanwhile, remains largely unconstrained but with several groups in hot pursuit, including those, for example, at NO$\nu$A \cite{nova2005}, T2K \cite{t2k2015}, and DUNE \cite{dune2015}.  Two additional phases are present if neutrinos are Majorana particles.  As the Majorana phases have no effect on oscillations, we take them to vanish.

The mismatch between the flavor and mass eigenstates is one of the fundamental facts about neutrinos.  It gives rise to oscillations in vacuum and is essential to the rich phenomenology of in-medium flavor evolution that has been discovered since Wolfenstein's pioneering revelation \cite{wolfenstein1978} that neutrinos propagating in matter sustain refractive effects in much the same way that photons do.  The derivation of neutrino oscillations and detection probabilities from $U_\textrm{PMNS}$ can be found in the standard references \cite{patrignani2016}.

Throughout this paper we confine our attention to the coherent limit of neutrino propagation, which is to say that collisions (scattering processes that alter the momentum of the neutrino) are negligible \cite{volpe2013, vlasenko2014, kartavtsev2015, keister2015}.  This approximation holds to varying degrees of accuracy in many settings of interest: It is applicable in vacuum, for one, as well as in astrophysical environments such as the Earth, the solar interior, the region far outside a core-collapse supernova or compact-object merger, and the early universe after weak decoupling, but it fails in the extremely dense interior of a supernova or merger remnant or at high enough temperatures in the early universe that neutrinos are thermally equilibrated with the plasma.  The coherent and incoherent limits are tied together by a regime in which neither collisions nor the medium-enhanced flavor transformation that occurs between scattering events can be neglected, as is the case during the protracted transition of neutrinos in the early universe from being strongly coupled to the plasma to being fully free-streaming.  This worst-of-both-worlds regime is also exemplified by the ``neutrino halo'' region of core-collapse supernovae \cite{cherry2012}, which is negligible during the late-time neutrino-driven-wind epoch but may be important during the neutronization burst or shock revival.  Environments bridging the coherent and incoherent extremes are a frontier of research in neutrino astrophysics and lie beyond the ambitions of the present paper.

In the coherent limit, which we henceforth adopt, the neutrino flavor state obeys a Schr\"odinger-like equation
\begin{equation}
i \frac{d | \psi \rangle}{dt} = H | \psi \rangle, \label{wfeom}
\end{equation}
with Hamiltonian
\begin{equation}
H = H_\textrm{vac} + H_\textrm{matt} + H_\nu,
\end{equation}
where the three contributions to the Hamiltonian are respectively due to neutrino mass, forward scattering off of matter particles (nucleons, charged leptons), and forward scattering off of other neutrinos \cite{sigl1993}.  We briefly discuss each of these in turn.

The vacuum term $H_\textrm{vac}$ is present in all environments and encodes the masses of the individual neutrino eigenstates.  In the mass basis it is simply the matrix $\left( 1 / 2 E \right) \textrm{diag} \left( m_1^2, m_2^2, m_3^2 \right)$, where $m_i$ is the mass of eigenstate $\nu _i$ and $E$ is the neutrino energy.  In the flavor basis it has the form
\begin{equation}
H_\textrm{vac} = U_\textrm{PMNS} \left( \frac{1}{2 E} \textrm{diag} \left( m_1^2, m_2^2, m_3^2 \right) \right) U^\dag_\textrm{PMNS}. \label{hvacgen}
\end{equation}
It is evident that $H_\textrm{vac}$ has non-zero off-diagonal elements in the flavor basis that cause mixing between the flavor states: This, of course, is the phenomenon of neutrino oscillations in vacuum.

When neutrinos are immersed in a dense bath of matter particles, as they are in many astrophysical settings, the dispersion relations of the individual flavors are modified by forward scattering off of the background.  The most common scenario is one in which electrons (and possibly positrons) are abundant but muons and tauons are all but absent; a thermal environment requires quite a high temperature for the heavier charged leptons to be plentiful.  Under these circumstances, all flavors feel a forward-scattering potential generated by the neutral-current weak interaction with $e^\pm$, but only $\nu_e$ feels the additional potential from the charged-current interaction.  This effect is encoded in $H_\textrm{matt}$.  In this paper we are not concerned with the precise form of the matter Hamiltonian (for reasons that will become evident momentarily), so to illustrate its structure we write down the matrix in the scenario where the matter background consists entirely of $e^-$ with number density $n_e$.  In the flavor basis,
\begin{equation}
H_\textrm{matt} = \sqrt{2} G_F n_e \textrm{diag} \left( 1, 0, 0 \right), \label{hmatt3}
\end{equation}
where $G_F$ is the Fermi constant.  Since $H_\textrm{matt}$ is diagonal in the flavor basis and $H_\textrm{vac}$ is diagonal in the mass basis, the energy eigenstates in medium differ from both the mass and the flavor eigenstates.  Under the right conditions, the adiabatic decrease in $n_e$ from very high down to vanishing density induces efficient conversion through the MSW mechanism. 

The final constituent of the Hamiltonian stems from forward scattering with other neutrinos.  The physics introduced by this term is rich, as it generalizes the effect of $H_\textrm{matt}$ to a nonlinear, matrix-structured index of refraction.  The matrix structure enters because the neutrino--neutrino forward-scattering amplitude depends not just on the density of the background neutrinos but on their quantum states.  The potential generated by these processes is therefore proportional to a sum over the density matrices $\rho = | \psi \rangle \langle \psi |$ of each background neutrino.  Explicitly,
\begin{equation}
\rho = \left( \begin{array}{ccc}
\rho_{ee} & \rho_{e\mu} & \rho_{e\tau} \\
\rho_{e\mu}^* & \rho_{\mu\mu} & \rho_{\mu\tau} \\
\rho_{e\tau}^* & \rho_{\mu\tau}^* & \rho_{\tau\tau}
\end{array} \right) = \left( \begin{array}{ccc}
| a_e |^2 & a_e^* a_\mu & a_e^* a_\tau \\
a_\mu^* a_e & | a_\mu |^2 & a_\mu^* a_\tau \\
a_\tau^* a_e & a_\tau^* a_\mu & | a_\tau |^2
\end{array} \right).
\end{equation}
The diagonal element $\rho_{\alpha\alpha}$ is proportional to the number density of neutrinos of flavor $\alpha$ and the off-diagonal element $\rho_{\alpha\beta}$ ($\alpha \neq \beta$) measures the quantum coherence between flavors $\alpha$ and $\beta$.  If neutrinos of momentum $\vec{q}$ have number density $n_{\nu, \vec{q}}$ and flavor state $\rho_{\vec{q}}$, then a neutrino of momentum $\vec{p}$ propagating through this background experiences
\begin{equation}
H_\nu = \sqrt{2} G_F \sum_{\vec{q}} \left( 1 - \hat{p} \cdot \hat{q} \right) n_{\nu, \vec{q}} \rho_{\vec{q}}, \label{selfpot}
\end{equation}
where the sum is over all momentum states but could be expanded to include any additional indices used to label neutrinos in the system.  (By writing $\rho = | \psi \rangle \langle \psi |$, we have assumed that each density matrix describes a pure state.)  The geometric factor $\left( 1 - \vec{p} \cdot \vec{q} \right)$ originates from the structure of the weak-interaction current.  For the sake of brevity, we will later use $\mu_{\vec{q}} \equiv \sqrt{2} G_F \left( 1 - \hat{p} \cdot \hat{q} \right) n_{\nu, \vec{q}}$.  Note that this contribution is nonlinear in the sense that it couples together the different neutrino trajectories in flavor space.  (Note also that we are ignoring antineutrinos in this discussion.  We will continue to do so in the calculations that follow, as antineutrinos do not change the analysis in any essential way.) 

Throughout much of this paper we will perform calculations in the two-flavor approximation that is appropriate when $\nu_\mu$ and $\nu_\tau$ have the same interaction potentials, which holds whenever (1) muons and tauons are scarce and (2) $\nu_\mu$ and $\nu_\tau$ have identical spectra.  Specifically, we consider mixing between $\nu_e$ and a state $\nu_x$, the latter being a particular superposition of $\nu_\mu$ and $\nu_\tau$.  In this case $| \psi \rangle$, $\rho$, and the interaction potentials reduce in an obvious manner from the three-flavor expressions given above.  There is now just a single mixing angle $\theta_\textrm{v}$, with no CP-violating phase, and the $2 \times 2$ mixing matrix is simply the rotation
\begin{equation}
U = \left( \begin{array}{cc}
\cos\theta_\textrm{v} & \sin\theta_\textrm{v} \\
- \sin\theta_\textrm{v} & \cos\theta_\textrm{v}
\end{array} \right).
\end{equation}
It follows from Eq.~\eqref{hvacgen} that the vacuum Hamiltonian is
\begin{equation}
H_\textrm{vac} = \frac{\omega}{2} \left( \begin{array}{cc}
- \cos 2\theta_\textrm{v} & \sin 2 \theta_\textrm{v} \\
\sin 2 \theta_\textrm{v} & \cos 2 \theta_\textrm{v}
\end{array} \right),
\end{equation}
with the vacuum oscillation frequency defined in terms of the mass-squared splitting $\delta m^2 \equiv m_2^2 - m_1^2$ by $\omega \equiv \delta m^2 / 2 E$.  The mass hierarchy, which remains experimentally ambiguous, is reflected in the sign of the oscillation frequency: $\omega > 0$ for the normal hierarchy (NH), $\omega < 0$ for the inverted hierarchy (IH).

With only two flavors, coherent neutrino evolution is tantamount to a two-level problem and can be mapped onto the Bloch sphere; in this regard it is analogous to the physics of electron spins, nuclear isospins, qubits, and so on.  The Bloch vector $\vec{P}$ formed from the $\mathrm{su}(2)$-algebra decomposition of the density matrix $\rho$ is commonly known in the neutrino community as the polarization vector, in deference to photon polarization:
\begin{equation}
\rho = \frac{1}{2} \left( \mathbb{I} + \vec{P} \cdot \vec{\sigma} \right). \label{vecpdef}
\end{equation}
Noting that Eq.~\eqref{wfeom} can be recast as a Liouville--von Neumann equation
\begin{equation}
i \frac{d \rho}{dt} = \left[ H, \rho \right],
\end{equation}
a similar decomposition of the Hamiltonian permits the coherent equations of motion to be written as a Bloch-like equation with infinite relaxation time:
\begin{equation}
\frac{d \vec{P}}{dt} = \vec{H} \times \vec{P}. \label{veceom}
\end{equation}
$\vec{P}$ can be visualized as a vector pointing from the origin to the surface of the $S^2$ manifold described at the end of the previous section.  In this picture the parallel transport condition in Eq.~\eqref{parallel} forbids $\vec{P}$ from spinning and thereby moving locally along the fiber, even as $\vec{P}$ precesses about $\vec{H}$.  Given our emphasis on the two-flavor limit, it will be helpful to have this polarization-vector picture in mind.  Indeed, it is precisely the geometric nature of Eq.~\eqref{veceom} that underlies the geometric phases exhibited below.

Absent the self-coupling potential $H_\nu$, the Hamiltonian can be rewritten in terms of effective in-medium mixing parameters:
\begin{equation}
H_\textrm{vac} + H_\textrm{matt} = \frac{\omega_\textrm{m}}{2} \left( \begin{array}{cc}
- \cos 2\theta_\textrm{m} & \sin 2 \theta_\textrm{m} \\
\sin 2 \theta_\textrm{m} & \cos 2 \theta_\textrm{m}
\end{array} \right),
\end{equation}
with in-medium oscillation frequency (using the analogous form of Eq.~\eqref{hmatt3} for two flavors)
\begin{equation}
\omega_\textrm{m} \equiv \sqrt{\omega \sin^2 2\theta_\textrm{v} + \left( \omega \cos 2\theta_\textrm{v} -  \frac{\sqrt{2}}{2} G_F n_e \right)^2 }
\end{equation}
and in-medium mixing angle given by
\begin{equation}
\sin^2 2\theta_\textrm{m} \equiv \frac{\omega^2 \sin^2 2 \theta_\textrm{v}}{\omega_\textrm{m}^2}.
\end{equation}
The upshot is that flavor evolution in a matter background looks like vacuum oscillations with modified frequency and amplitude.  For this reason, in the rest of the paper we will ignore $H_\textrm{matt}$; it is assumed to have been absorbed into the vacuum mixing parameters.

With $H_\nu$ present, the nonlinear communication between flavor states considerably expands the range of flavor-evolution phenomena.  These behaviors are grouped under the heading of \textit{collective neutrino oscillations} and have been the subject of intense study in recent years \cite{kostelecky1993, kostelecky1993b, samuel1993, kostelecky1994, kostelecky1995, qian1995a, balantekin1999, pastor2002, pastor2002b, bell2003, balantekin2005, duan2006, duan2006b, duan2006c, duan2006d, hannestad2006, balantekin2007, duan2007, duan2007b, duan2007c, fogli2007, raffelt2007, raffelt2007b, raffelt2007c, blennow2008, chakraborty2008, dasgupta2008, dasgupta2008b, dasgupta2008c, duan2008, duan2008b, duan2008c, esteban2008, esteban2008b, gava2008, dasgupta2009, duan2009, duan2009b, fogli2009, gava2009, sawyer2009, friedland2010, duan2010, chakraborty2010, dasgupta2010, banerjee2011, chakraborty2011, chakraborty2011b, duan2011, galais2011, pehlivan2011, raffelt2011, dasgupta2012, gouvea2012, sarikas2012, sarikas2012b, raffelt2013, raffelt2013b, vlasenko2014b, chakraborty2016, malkus2016, johns2016, tian2016, armstrong2016}.  One paradigmatic collective effect is the synchronization of flavor: When $H_\nu$ dominates the Hamiltonian, all neutrinos experience roughly the same potential, leading them to undergo nearly identical oscillations at a common effective frequency.  Synchronized oscillations are perhaps the cleanest example of cyclic evolution of the Hamiltonian at strong nonlinear coupling, but they are not alone.

Geometrically the crucial feature of neutrino self-coupling is that even in two flavors a complex off-diagonal potential can develop, opening the possibility for cyclic evolution of the Hamiltonian.  With the standard matrix representation of the Pauli matrices, the $y$-component of $\vec{H}$ corresponds to the imaginary parts of these off-diagonal elements.  Since $H$ can always be written as a real symmetric matrix in the standard MSW (vacuum+matter) scenario, in a matter background $\vec{H}$ never leaves the $xz$-plane and closed loops on the Bloch sphere, other than the trivial one formed by following the $xz$ great circle, are precluded.  This conclusion no longer holds in a \textit{neutrino} background.  The coherent coupling of neutrino flavor states can thus be framed as a geometric statement.

\section{Mixed potentials with two flavors \label{2mixedsec}}

In this and the next two sections we analyze geometric effects that arise in a scenario with two-flavor neutrinos interacting with each other by way of coherent forward scattering.  For simplicity we neglect any contributions from a matter background, which as noted previously can be absorbed into the vacuum potential by working in terms of effective in-medium mixing parameters.  By dialing the strengths of the vacuum and self-coupling potentials, one finds that the system gives rise to a panoply of flavor-transformation phenomena.  We take three different limits of the coupling strengths that illuminate in particular how the flavor transformation enmeshes with geometry.

The equations of motion for two pure populations of neutrinos interacting with one another are
\begin{align}
& i \frac{d | \psi_1 (t) \rangle}{dt} = \left[ \omega_1 B + \mu_2 \rho_2 (t) \right] | \psi_1 (t) \rangle, \notag \\
& i \frac{d | \psi_2 (t) \rangle}{dt} = \left[ \omega_2 B + \mu_1 \rho_1 (t) \right] | \psi_2 (t) \rangle, \label{starteom}
\end{align}
where mode $i$ has wavefunction $| \psi_i (t) \rangle$, vacuum oscillation frequency $\omega_i$, and density parameter $\mu_i$ (defined below Eq.~\eqref{selfpot}).  The matrix $B$ is equal to $H_\textrm{vac}$ with the energy-dependent part taken out.  In the mass basis it is $B = \textrm{diag} \left( -1/2, 1/2 \right)$ and has vector form $\vec{B} = - \left( 1 / 2 \right) \hat{z}$.

In this section we consider the limit in which the neutrinos of mode 2 are extremely dilute but those of mode 1 are extremely dense:
\begin{align}
& i \frac{d | \psi_1 (t) \rangle}{dt} = \omega_1 B | \psi_1 (t) \rangle, \notag \\
& i \frac{d | \psi_2 (t) \rangle}{dt} = \mu_1 \rho_1 (t) | \psi_2 (t) \rangle.
\end{align}
As a shorthand, we term this arrangement \textit{mixed potentials}.

The first equation of motion describes vacuum oscillations and is easily solved:
\begin{equation}
| \psi_1 (t) \rangle = \exp \left( - i \omega_1 B t \right) | \psi_1 (0) \rangle.
\end{equation}
In the mass basis the matrix exponential is diagonal and, taking as an initial state
\begin{equation}
| \psi_1 (0) \rangle = \left( \begin{array}{c}
\cos \frac{\theta_1}{2} \\
e^{i \phi_1} \sin \frac{\theta_1}{2}
\end{array} \right),
\end{equation}
one finds that $| \psi_1 \rangle$ corresponds to a polarization vector $\vec{P}_1$ precessing about the $z$-axis with fixed frequency $\omega_1$ and at fixed polar angle $\theta_1$:
\begin{equation}
| \psi_1 (t) \rangle = \left( \begin{array}{c}
\cos \frac{\theta_1}{2} \\
e^{i ( \phi_1 - \omega_1 t)} \sin \frac{\theta_1}{2}
\end{array} \right).
\end{equation}
In the case of the IH, $\omega_1 < 0$ and the direction of precession is reversed.  The other mode $| \psi_2 \rangle$---a flavor state evolving under a Hamiltonian that sweeps out a circle in flavor space---is mathematically identical to a spin in a magnetic field that sweeps out a circle in \textit{physical} space.  Although we are indicating with the notation that $| \psi_1 \rangle$ represents a pure state of neutrinos at a chosen energy, it may be that $| \psi_2 \rangle$ interacts with an ensemble---pure or mixed---with some spectrum.  If the ensemble undergoes synchronized oscillations, then the computation proceeds almost unchanged.

In Sec.~\ref{mixedadiab} we adopt an adiabatic treatment, thereby reproducing the neutrino version of the classic result for the geometric phase of a spin in a cyclic magnetic field, and we point out that in principle this phase is observable.  In Sec.~\ref{mixedexact} we find the exact (non-adiabatic) solution and demonstrate the geometric-dynamical phases that appear as perturbative corrections to the traditional purely geometric phase.

\subsection{Adiabatic treatment \label{mixedadiab}}

\begin{figure}
\includegraphics[width=.48\textwidth]{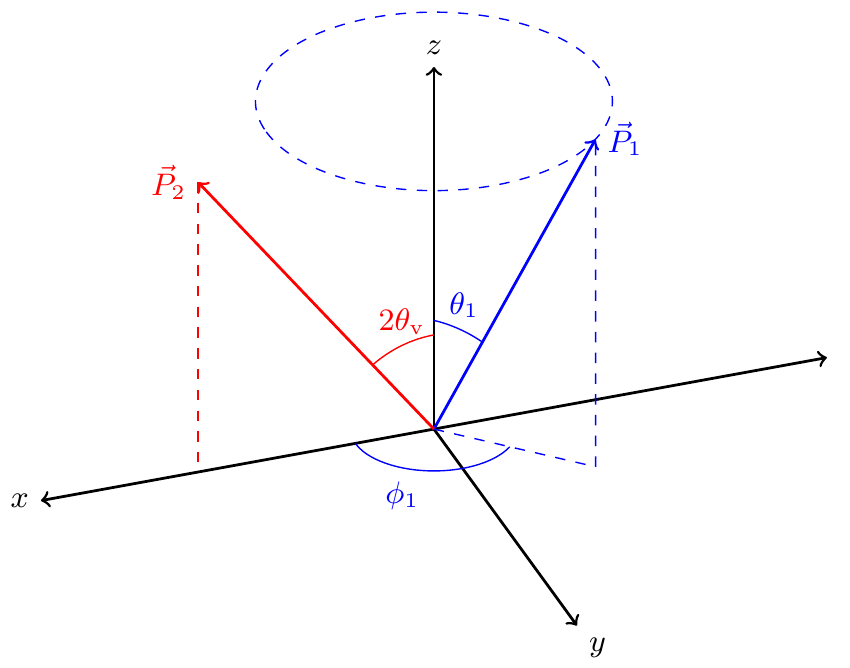}
\caption{Initial ($t = 0$) configuration of polarization vectors in the mixed-potentials scenario. $\vec{P}_1$ undergoes vacuum oscillations (clockwise about $\hat{z} = - \hat{B}$ for the NH, counterclockwise for the IH). Its trajectory is shown by the dashed circle.  $\vec{P}_2$ is in an electron-flavor eigenstate and points along $\hat{L}$.  \label{mixedfig}}
\end{figure}

We now set out to determine, under the assumption of adiabatic evolution, the phase acquired by $| \psi_2 \rangle$ after $\vec{H}_2 (t) = \mu_1 \vec{P}_1 (t)$ undergoes one period of cyclic evolution.  Based on the foregoing discussion, we know that $\vec{H}_2$ rotates with frequency $\omega_1$ about the mass-eigenstate axis $\hat{B} = - \hat{z}$ ($\Phi_1 ( t ) =  \phi_1 - \omega_1 t$), maintaining a constant magnitude $|\vec{H}_2 (t) | = \mu_1$ and a constant polar angle $\Theta_1 (t) = \theta_1$.  The coordinate system is chosen such that the flavor-eigenstate axis $\hat{L}$, which is defined to point along the polarization vector associated with $\nu_e$, is in the $xz$-plane.  This set-up is depicted in Fig.~\ref{mixedfig}.

%\begin{figure}
%\tdplotsetmaincoords{60}{160}
%\begin{tikzpicture}[scale=.85,tdplot_main_coords]
%
%\draw[thick,->] (0,0,0) -- (5,0,0) node[anchor=east]{$x$};
%\draw[thick,->] (0,0,0) -- (-5,0,0) node[anchor=west]{};
%\draw[thick,->] (0,0,0) -- (0,5,0) node[anchor=north west]{$y$};
%\draw[thick,->] (0,0,0) -- (0,0,5) node[anchor=south]{$z$};
%
%\pgfmathsetmacro{\ax}{-1.508}
%\pgfmathsetmacro{\ay}{1.508}
%\pgfmathsetmacro{\az}{4.523}
%
%\draw[thick,->,blue] (0,0,0) -- (\ax,\ay,\az) node[anchor=west]{$\vec{P}_1$};
%\draw[thick,->,red](0,0,0) -- (3,0,4) node[anchor=east]{$\vec{P}_2$};
%\draw[dashed,blue] (0,0,0) -- (\ax,\ay,0) -- (\ax,\ay,\az);
%\draw[dashed,red] (3,0,4) -- (3,0,0);
%
%\tdplotgetpolarcoords{\ax}{\ay}{\az}
%\tdplotsetthetaplanecoords{\tdplotresphi}
%\tdplotdrawarc[blue]{(0,0,0)}{1}{0}%
%{\tdplotresphi}{anchor=north east}{$\phi_1$}
%\tdplotdrawarc[blue,dashed]{(0,0,4.523)}{2.133}{0}%
%{360}{anchor=west}{}
%\tdplotdrawarc[tdplot_rotated_coords,blue]{(0,0,0)}{1.5}{0}%
%{\tdplotrestheta}{anchor=south}{~$\theta_1$}
%
%\tdplotgetpolarcoords{3}{0}{4}
%\tdplotsetthetaplanecoords{\tdplotresphi}
%\tdplotdrawarc[tdplot_rotated_coords,red]{(0,0,0)}{1.3}{0}%
%{\tdplotrestheta}{anchor=south}{$2\theta_\textrm{v}$}
%
%\end{tikzpicture}
%\caption{Initial ($t = 0$) configuration of polarization vectors in the mixed-potentials scenario. $\vec{P}_1$ undergoes vacuum oscillations (clockwise about $\hat{z} = - \hat{B}$ for the NH, counterclockwise for the IH). Its trajectory is shown by the dashed circle.  $\vec{P}_2$ is in an electron-flavor eigenstate and points along $\hat{L}$. \label{mixedfig}}
%\end{figure}

If the test neutrino is initially electron flavor, that is,
\begin{equation}
\left| \psi_2 ( 0 ) \right\rangle = \left| \nu_e \right\rangle,
\end{equation}
then $\vec{P}_2$ has initial azimuthal angle $\Phi_2 (0) = 0$ and initial polar angle $\Theta_2 (0) = 2 \theta_\textrm{v}$, where $\theta_\textrm{v}$ is the mixing angle in vacuum.  We stress that this problem has four physically relevant vectors.  Making the reasonable stipulation that $| \psi_1 \rangle$ also decouples into a flavor eigenstate, then at any given time the vectors $\hat{L}$, $\hat{B}$, $\hat{P}_1$, and $\hat{P}_2$ are generally not coincident.  One can see quite readily that should any two of these unit vectors be identical at all times, then geometric phases in $| \psi_2 \rangle$ are either absent or unobservable:
\begin{itemize}
\item If $\hat{B} = \hat{L}$, then oscillations do not occur. 
\item If $\hat{P}_1 (t) = \hat{B}$, then the path of $\vec{H}_2$ does not enclose a finite area in parameter space.
\item If $\hat{P}_2 (t) = \hat{P}_1 (t)$, then in the adiabatic limit $| \psi_2 \rangle$ is in an energy eigenstate at all times and its phase will not show up at the probability level.
\item If $\hat{P}_1 (t) = \hat{L}$, $\hat{P}_2 (t) = \hat{L}$, or $\hat{P}_2 (t) = \hat{B}$, then $\hat{L}$ must be equal to $\hat{B}$.
\end{itemize}
It follows immediately that if decoupling occurs into flavor eigenstates, it is a prerequisite for the appearance of an observable adiabatic geometric phase that two parameters be nonzero: the initial relative phase $\phi_1$ between the two modes and the vacuum mixing angle $\theta_\textrm{v}$.  With these considerations in mind, we now proceed to derive the geometric phase.

In the chosen coordinate system the Hamiltonian matrix is
\begin{equation}
H_2 = \frac{\mu_1}{2} \left( \begin{array}{cc}
\cos\theta_1 & e^{-i \left( \phi_1 - \omega_1 t \right)} \sin\theta_1 \\
e^{i \left( \phi_1 - \omega_1 t \right)} \sin\theta_1 & -\cos\theta_1
\end{array} \right),
\end{equation}
which is represented by the vector
\begin{equation}
\vec{H}_2 = \frac{\mu_1}{2} \left( \begin{array}{c}
\sin\theta_1 \cos \left( \phi_1 - \omega_1 t \right)  \\
\sin\theta_1 \sin \left( \phi_1 - \omega_1 t \right) \\
\cos\theta_1
\end{array} \right).
\end{equation}
The energy eigenstates of this Hamiltonian correspond to the normalized polarization vectors parallel and antiparallel with $\vec{H}_2$.  As kets they are
\begin{align}
\left| \nu _+ (t) \right\rangle = \left( \begin{array}{c}
\cos\frac{\theta_1}{2} \\
e^{i \left( \phi_1 - \omega_1 t \right)} \sin\frac{\theta_1}{2}
\end{array} \right), \notag \\
\left| \nu _- (t) \right\rangle = \left( \begin{array}{c}
-\sin\frac{\theta_1}{2} \\
e^{i \left( \phi_1 - \omega_1 t \right)} \cos\frac{\theta_1}{2}
\end{array} \right),
\end{align}
and they have energy eigenvalues $E_\pm = \pm \mu_1 / 2$.

The geometric phase, defined in Eq.~\eqref{gammadef}, can be recast in the form
\begin{equation}
\gamma = \oint _\mathcal{C} \vec{A} \cdot d\vec{R}, ~~~~~ \vec{A} = i \langle \eta (t) | \nabla | \eta (t) \rangle
\end{equation}
for eigenstate $| \eta (t) \rangle$.  The vector $\vec{A}$ is the gauge potential associated with the adiabatic connection.  In this case the gauge potentials are given by
\begin{align}
\vec{A}_+ = i \left\langle \nu _+ \right| \nabla \left| \nu _+ \right\rangle = - \frac{\sin ^2 \frac{\Theta_1}{2}}{\sin\Theta_1} \hat{\Phi}, \notag \\
\vec{A}_- = i \left\langle \nu _- \right| \nabla \left| \nu _- \right\rangle = - \frac{\cos ^2 \frac{\Theta_1}{2}}{\sin\Theta_1} \hat{\Phi}.
\end{align}
Integration along the curve swept out by $\vec{H}_2$ then yields the geometric phases acquired by these energy eigenstates:
\begin{align}
\gamma_+ = \oint_C \vec{A}_+ \cdot d\vec{R} = \pm \pi \left( 1 - \cos\theta_1 \right), \notag \\
\gamma_- = \oint_C \vec{A}_- \cdot d\vec{R} = \pm \pi \left( 1 + \cos\theta_1 \right),
\end{align}
where the upper (lower) signs are for the NH (IH).  Evidently the geometric phase is sensitive to the mass hierarchy, with the proper sign being fixed by the direction of traversal about the loop.  The dynamical phase, meanwhile, is
\begin{equation}
\delta_\pm = - \int_0 ^T E_\pm dt = \mp \frac{\mu_1}{2} T,
\end{equation}
so that the energy eigenstates after one period $T$ are
\begin{align}
& \left| \nu _+ ( T ) \right\rangle = e^{\pm i \pi \left( 1 - \cos\theta_1 \right)} e^{- i \frac{\mu_1}{2} T} \left| \nu _+ ( 0 ) \right\rangle, \notag \\
& \left| \nu _- ( T ) \right\rangle = e^{\pm i \pi \left( 1 + \cos\theta_1 \right)} e^{+ i \frac{\mu_1}{2} T} \left| \nu _- ( 0 ) \right\rangle,\end{align}
where again the upper (lower) signs are for the NH (IH).

The calculation thus far is identical to the standard one for a spin-$1/2$ particle in a rotating magnetic field, and as usual the geometric phase is half the solid angle enclosed in parameter space by the loop traced out by the Hamiltonian vector.  But a key point for neutrinos is that their production and detection project onto the flavor axis.  It is therefore necessary to convert between the interaction and energy bases.  The unitary matrix $\mathcal{U}$ effecting the transformation
\begin{equation}
\left( \begin{array}{c}
\left| \nu _- \right\rangle \\
\left| \nu _+ \right\rangle
\end{array} \right)
= \mathcal{U}
\left( \begin{array}{c}
\left| \nu _e \right\rangle \\
\left| \nu _x \right\rangle
\end{array} \right)
\end{equation}
is given by
\begin{widetext}
\begin{equation}
\mathcal{U} = \left( \begin{array}{cc}
\mathcal{U}_{11} & \mathcal{U}_{12} \\
\mathcal{U}_{21} & \mathcal{U}_{22}
\end{array} \right)
=
\left( \begin{array}{cc}
e^{i \Phi_1} \sin\theta_\textrm{v}\cos\frac{\theta_1}{2} - \cos\theta_\textrm{v} \sin\frac{\theta_1}{2} & e^{i \Phi_1} \cos\theta_\textrm{v} \cos \frac{\theta_1}{2} + \sin\theta_\textrm{v}\sin\frac{\theta_1}{2} \\
e^{i \Phi_1} \sin\theta_\textrm{v}\sin\frac{\theta_1}{2} + \cos\theta_\textrm{v} \cos\frac{\theta_1}{2} & e^{i \Phi_1} \cos\theta_\textrm{v} \sin\frac{\theta_1}{2} - \sin\theta_\textrm{v}\cos \frac{\theta_1}{2}
\end{array} \right).
\end{equation}
\end{widetext}
Note that $\mathcal{U}$ is time-dependent, since $\Phi_1 ( t ) = \phi_1 - \omega_1 t$.

Using the unitarity of $\mathcal{U}$, one can write the initial state---assumed to be $\nu_e$---as
\begin{equation} 
\left| \psi_2 ( 0 ) \right\rangle =  \mathcal{U}_{11}^* \left| \nu _- ( 0 ) \right\rangle + \mathcal{U}_{21}^* \left| \nu _+ ( 0 ) \right\rangle .
\end{equation}
After one period has elapsed, the state has evolved to
\begin{align}
\left| \psi_2 ( T ) \right\rangle & = \mathcal{U}_{11}^* e^{-i \pi \left(1 + \cos\theta_1 \right)} e^{i \frac{\mu_1}{2} T} \left| \nu _- (0) \right\rangle \notag \\
& ~~~ + \mathcal{U}_{21}^* e^{-i \pi \left(1 - \cos\theta_1 \right)} e^{-i \frac{\mu_1}{2} T} \left| \nu _+ ( 0 ) \right\rangle .
\end{align}
Projecting $\left| \psi_2 ( T ) \right\rangle$ onto the flavor state in which it was produced at $t = 0$ yields
\begin{equation}
\left| \left\langle \nu _e | \psi_2 ( T ) \right\rangle \right| ^2 = 1 - 4 \left| \mathcal{U}_{11} \right| ^2 \left| \mathcal{U}_{21} \right| ^2 \sin^2 \left( \pi \cos\theta_1 - \frac{\mu_1}{2} T \right).
\end{equation} 
Letting $x \equiv \cos 2\theta_\textrm{v} \cos\theta_1 + \cos\phi_1 \sin 2\theta_\textrm{v} \sin\theta_1$, we have
\begin{equation}
\left| \mathcal{U}_{11} \right| ^2 = \frac{1}{2} \left( 1 - x \right), ~~~ \left| \mathcal{U}_{21} \right| ^2 = \frac{1}{2} \left( 1 + x \right),
\end{equation}
allowing us to write
\begin{equation} 
\left| \left\langle \nu _e | \psi_2 ( T ) \right\rangle \right| ^2 = 1 - \left( 1 - x^2 \right) \sin^2 \left( \pi \cos\theta_1 - \frac{\mu_1}{2} T \right).
\end{equation} 
Recalling that $\gamma_\pm = - \pi \pm \pi \cos\theta_1$, one has, finally,
\begin{equation}
\left| \left\langle \nu _e | \psi_2 ( T ) \right\rangle \right| ^2 = 1 - \left( 1 - x^2 \right) \sin^2 \left( \gamma_\pm + \delta_\pm \right), \label{adiabresult}
\end{equation}
where the choice of $\pm$ is arbitrary.  We would have arrived at the same expression if we had instead chosen the neutrino to be initially $\nu_x$.

Moreover, Eq.~\eqref{adiabresult} is independent of the choice of hierarchy.  But since the overall sign of $\gamma$ changes upon flipping the hierarchy---whereas the sign of $\delta$ goes unchanged---the transition probability turns out to be hierarchy-dependent.  This finding has a simple explanation in the polarization-vector picture: The precession direction of $\vec{P}_1$ about $\hat{B}$ is set by the hierarchy, while the precession direction of $\vec{P}_2$ about $\hat{P}_1$ is not.

It may be helpful to note that $\left| \left\langle \nu _\pm ( 0 ) | \nu _\pm ( T ) \right\rangle \right| = 1$, since the geometric and dynamical phases vanish under the modulus.  In other words, the fact that the neutrino is produced and detected in a state other than one of the energy eigenstates is necessary for the phases to appear at the probability level.  In fact, if one knows the flavor of the neutrino at $t = 0$, then by measuring the flavor of the neutrino at $t = T$, one is effectively performing an interferometry experiment capable in principle of probing the geometric phase.  In this case, that phase is a measure of the flavor-space path traced out by the \textit{other} neutrino, which need not be directly observed.

%The probability-level phase revealed in this calculation is mathematically similar to that measured, for example, in experiments on photons traveling through a helical waveguide.  But in those experiments the apparatus must be designed to interfere photons that have traveled on two different paths \textbf{[true?]}.  In our case here the interference requires only one physical path, since a single path in real space corresponds to two paths in flavor space; the appearance of the phase therefore necessitates only detection.

\subsection{Exact solution \label{mixedexact}}

The formulae applied in the previous section are appropriate to the adiabatic limit, in which the energy eigenvectors track the Hamiltonian vector as it sweeps out a circuit.  But it turns out that an exact solution can be found even without this assumption.  Let
\begin{equation}
| \psi_2 (t) \rangle = a(t) | \nu _+ (t) \rangle + b (t) | \nu _- (t) \rangle
\end{equation}
and suppose that $| \psi_2 (0) \rangle = | \nu _+ (t) \rangle$, so that $a(0) = 1$ and $b(0) = 0$.  This initial condition is equivalent to the one in the previous subsection, but here we are not demanding that $| \psi_2 \rangle$ remain in the eigenstate $| \nu_+ \rangle$.

The Schr\"odinger equation says that the coefficients of $| \psi_2 (t) \rangle$ obey the system of equations
\begin{align}
\frac{da(t)}{dt} + a(t) \bigg\langle \nu _+ (t) \bigg| \frac{d}{dt} \bigg| \nu _+ (t) \bigg\rangle  + b(t) & \bigg\langle \nu _+ (t) \bigg| \frac{d}{dt} \bigg| \nu _- (t) \bigg\rangle \notag \\
& = - i \frac{\mu_1}{2} a(t), \notag \\
\frac{db(t)}{dt} + a(t) \bigg\langle \nu _- (t) \bigg| \frac{d}{dt} \bigg| \nu _+ (t) \bigg\rangle  + b(t) & \bigg\langle \nu _- (t) \bigg| \frac{d}{dt} \bigg| \nu _- (t) \bigg\rangle \notag \\
& = i \frac{\mu_1}{2} b(t).
\end{align}
Enforcing $b(t) = 0$ amounts to the adiabatic approximation; it can be seen that the deviation from this limit is associated with the ``cross terms'' that mix the eigenstates.  The coupled first-order differential equations can be rewritten as decoupled second-order differential equations.  The equation for $a(t)$ is
\begin{equation}
\frac{d^2 a}{dt^2} - i \omega_1 \frac{da}{dt} + \left[ \left( \frac{\mu_1}{2} \right) ^2 + \frac{\mu_1}{2} \omega_1 \cos\theta_1 \right] a = 0.
\end{equation}
This is the equation of a (complex) damped harmonic oscillator with real frequency-squared and imaginary friction and can be solved with the usual ansatz $a(t) \sim \exp \alpha t$.  The resulting algebraic equation for $\alpha$ has solutions
\begin{equation}
\alpha _\pm = \frac{i \omega_1}{2} \pm \frac{i \mu_1}{2} \sqrt{1 + 2 \frac{\omega_1}{\mu_1} \cos\theta_1 + \left( \frac{\omega_1}{\mu_1} \right)^2 }.
\end{equation}
The general solution, of course, can be written as 
\begin{equation}
a(t) = c_+ e^{\alpha_+ t} + c_- e^{\alpha _- t},
\end{equation}
and the initial condition $| \nu (0) \rangle = | \nu _+ (0) \rangle$ implies that
\begin{align}
& c_+ = \frac{1}{2} - \frac{1}{2 \Delta}\left[ 1 + \frac{\omega_1}{\mu_1} \cos\theta_1 \right], \notag \\
& c_- = \frac{1}{2} + \frac{1}{2 \Delta}\left[ 1 + \frac{\omega_1}{\mu_1} \cos\theta_1 \right],
\end{align}
where $\Delta$ is the square root of the discriminant,
\begin{equation}
\Delta \equiv \sqrt{ 1 + 2 \frac{\omega_1}{\mu_1} \cos\theta_1 + \left( \frac{\omega_1}{\mu_1} \right)^2 }.
\end{equation}
Observe that $\alpha_\pm$ are purely imaginary regardless of the values of $\omega_1$, $\mu_1$, and $\theta_1$.

An important quantity found throughout the neutrino literature is the adiabaticity parameter $\Upsilon$ (usually denoted $\gamma$, but our hands are tied), upon which the transition probability $P$ through a resonance depends exponentially: $P \approx e^{- \pi \Upsilon / 2}$ \cite{haxton1986}.  The parameter can be cast into the form \cite{johns2016}
\begin{equation}
\Upsilon \approx \frac{\left| H_T \right|^2}{\left| \dot{H}_z \right|}, \label{upsilondef}
\end{equation}
where $H_T = H_x^2 + H_y^2$ is the transverse part of the Hamiltonian vector and the right-hand side is evaluated at resonance.  In circumstances where a flavor-state level crossing occurs, such as in the MSW mechanism, this definition implies that transitions are unlikely to occur if the separation between the energy eigenstates at closest approach is large relative to the speed with which the resonance is traversed.  Although the system we are analyzing has no such level crossing, $\Upsilon$ nonetheless coheres with what we mean by adiabaticity.  Applying the definition above, one has
\begin{equation}
\Upsilon \approx \frac{\mu_1 \left| P_{1, T} \right|^2}{\left| \omega_1 \left( \vec{B} \times \vec{P}_1 \right)_z \right|}.
\end{equation}
Dropping factors of order unity, this becomes simply $\Upsilon \approx \mu_1 / \omega_1$, so that the adiabatic limit corresponds to $\omega_1 / \mu_1 \longrightarrow 0$.  Thus the neutrino adiabaticity parameter, even in this non-resonant scenario, is consistent with the more general intuition that adiabaticity prevails when the change in the Hamiltonian is slow compared to the response of the particle.

To zeroth order in $\omega_1 / \mu_1$ one has, for $T = 2 \pi / \omega_1$,
\begin{equation}
\alpha_\pm T \longrightarrow \pm i \frac{\mu_1}{2} T + i \pi \left( 1 \mp \cos\theta_1 \right).
\end{equation}
The dynamical and geometric phases from the previous section are therefore recovered as the leading-order terms in the perturbation expansion in the adiabaticity parameter.

The probability of $| \psi_2 (t) \rangle$ being in the upper eigenstate at any time $t$ is
\begin{equation}
| a(t) |^2 = 1 - 2 c_+ c_- \left( 1 - \cos \mu_1 \Delta t \right).
\end{equation}
It is interesting to coerce $a(t)$ into the form $r(t) \exp i \phi (t)$.  The modulus is simply $r(t) = | a(t) |$ and the phase is
\begin{align}
\phi &(t) = \notag \\
& \arctan \left[ \frac{c_+ \sin \left( \frac{\omega_1 t}{2} + \frac{\mu_1 \Delta t}{2} \right) + c_- \sin \left( \frac{\omega_1 t}{2} - \frac{\mu_1 \Delta t}{2} \right)}{c_+ \cos \left( \frac{\omega_1 t}{2} + \frac{\mu_1 \Delta t}{2} \right) + c_- \cos \left( \frac{\omega_1 t}{2} - \frac{\mu_1 \Delta t}{2} \right)} \right].
\end{align}
Specifying $t = T$ leads to
\begin{align}
& r(T) = \sqrt{1-2c_+c_-\left( 1 - \cos \mu_1 \Delta T \right)}  \notag \\
& \phi (T) = \arctan \left[ \left( 1 - 2 c_- \right) \tan \frac{\mu_1 \Delta T}{2} \right],
\end{align}
Expanding each to first order in $\omega_1 / \mu_1$ yields
\begin{align}
r(T) \approx & ~1 - \frac{\omega_1}{2 \mu_1} \sin ^2 \theta_1 \sin^2 \left( \frac{\mu_1}{2} T + \pi \cos\theta_1 \right), \notag \\
\phi (T) \approx & - \frac{\mu_1}{2} T - \pi \cos\theta_1 - \frac{\omega_1}{\mu_1} \frac{\pi \sin ^2 \theta_1}{2}.
\end{align}
These can be combined to give an expression for $a(T)$, with the adiabatic-limit geometric and dynamical phases substituted appropriately:
\begin{align}
a(T) \approx & -  \left[ 1 + \frac{\pi}{2 \delta_+} \sin^2 \theta_1 \sin^2 \left( \delta_+ + \gamma_+ \right) \right] \notag \\
& \times \exp \left\lbrace i \left[ \delta_+ + \gamma_+ + \frac{\pi^2 \sin^2 \theta_1}{2 \delta_+} \right] \right\rbrace.
\end{align}
As expected, in the zeroth-order expansion one obtains
\begin{equation}
a(T) = - \exp \left[ i \left( \delta_+ + \gamma_+ \right) \right].
\end{equation}
This is identical to our result from the previous subsection, up to an unobservable minus sign.  Note also that the corrections to the fully adiabatic result intertwine geometry and dynamics.  It is only at lowest order that the two can be neatly separated.

\section{Pure self-coupling with two flavors \label{puresec}}

We have seen that if the Hamiltonian for a neutrino sweeps out a circle, then the neutrino acquires a geometric phase after one period that is proportional to the solid angle of this circle on the Bloch sphere.  It is well-known from geometric-phase lore that in fact the path could be any closed circuit and in all cases the phase acquired is determined by the enclosed solid angle.

In the neutrino context with strong nonlinear coupling between modes, the possibility arises that the geometric phase is not \textit{set} by the path but rather that the phase and the path mutually determine one another.  The simplest case, which we shall examine in this section, is that of two modes interacting with one another and experiencing negligible vacuum potential.  We can picture this scenario as two vectors rotating about each other in some complicated way.  If it can be shown that the Hamiltonian generated by one vector $\vec{P}_1$ is cyclic (\textit{i.e.}, if that vector is itself cyclic) and if the other vector $\vec{P}_2$ does not adiabatically track an energy eigenstate, then it is to be expected that geometric phases will appear at the probability level in the second mode, which is to say that the position of $\vec{P}_2$ depends on the geometric phases generated by $\vec{P}_1$.  Thus far all of this applies equally to the mixed-potentials scenario, as we just saw.  But with two neutrino populations interacting solely through self-coupling, these considerations are mutual, implying that the paths and geometric phases of the vectors are inextricably bound.  The scenario of the previous section was analogous to a spin in a rotating magnetic field; the scenario here is more akin to two spins interacting through their magnetic moments.

With these thoughts in mind, we return to the general equations of motion in Eq.~\eqref{starteom} and set $\omega_1 = \omega_2 = 0$, leaving the self-coupling potentials nonzero.  To be explicit, we have
\begin{align}
& i \frac{d | \psi_1 \rangle}{dt} = \mu_2 \rho_2 | \psi_1 \rangle, \notag \\
& i \frac{d | \psi_2 \rangle}{dt} = \mu_1 \rho_1 | \psi_2 \rangle. \label{pureeom}
\end{align}
Formally the solutions are
\begin{align}
& | \psi_1 (t) \rangle = \mathcal{P} \exp \left( -i \mu_2 \int_0^t dt' \rho_2(t') \right) | \psi_1 (0) \rangle, \notag \\
& | \psi_2 (t) \rangle = \mathcal{P} \exp \left( -i \mu_1 \int_0^t dt' \rho_1(t') \right) | \psi_2 (0) \rangle,
\end{align}
where $\mathcal{P}$ denotes the path-ordering operator, but clearly these expressions are of little help since the equations have not been decoupled.

In fact the equations can be decoupled, allowing for exact solutions to be obtained.  First note the important fact that
\begin{equation}
i \frac{d}{dt} \langle \psi_2 | \psi_1 \rangle = \left( \mu_2 - \mu_1 \right) \langle \psi_2 | \psi_1 \rangle, 
\end{equation}
hence the solution at time $t$ is given by
\begin{equation}
\langle \psi_2 (t) | \psi_1 (t) \rangle = \exp \left[ -i \left( \mu_2 - \mu_1 \right) t \right] \langle \psi_2 (0) | \psi_1 (0) \rangle \label{12overlap}
\end{equation}
and $| \langle \psi_2 (t) | \psi_1 (t) \rangle |^2$ is constant.  The geometric meaning of these statements is more transparent when Eq.~\eqref{pureeom} is rephrased in terms of polarization vectors:
\begin{align}
& \frac{d \vec{P}_1}{dt} = \mu_2 \vec{P}_2 \times \vec{P}_1, \notag \\
& \frac{d \vec{P}_2}{dt} = \mu_1 \vec{P}_1 \times \vec{P}_2.
\end{align}
The magnitudes of the polarization vectors are conserved as usual, as is $\vec{P}_1 \cdot \vec{P}_2$, and the conservation of $| \langle \psi_2 (t) | \psi_1 (t) \rangle |^2$ corresponds to the preservation of the angle between $\vec{P}_1$ and $\vec{P}_2$ even as the vectors drift through flavor space.  From a certain viewpoint, these are consequence of the conservation of $\vec{D} \equiv \mu_1 \vec{P}_1 + \mu_2 \vec{P}_2$, which acts as a kind of ``center of flavor'' in analogy to the center of mass of a mechanical system.

The first equation of motion in Eq.~\eqref{pureeom} can be rearranged to read
\begin{equation}
| \psi_2 \rangle = \frac{1}{\mu_2 \langle \psi_2 | \psi_1 \rangle} i \frac{d | \psi_1 \rangle}{dt}.
\end{equation}
Differentiating this---while keeping in mind Eq.~\eqref{12overlap}---and using the second equation of motion yields
\begin{equation}
\frac{d^2 | \psi_1 \rangle}{dt^2} + i \left(\mu_2 - \mu_1 \right) \frac{d | \psi_1 \rangle}{dt} + \mu_1 \mu_2 | \langle \psi_2 | \psi_1 \rangle |^2 | \psi_1 \rangle = 0. \label{pureosc}
\end{equation}
$| \psi_1 \rangle$ obeys the complex conjugate of this equation.  As with the decoupled equations of motions in the mixed-potentials limit, the friction coefficient is imaginary and the frequency-squared is real.

\begin{figure}
\includegraphics[width=.42\textwidth]{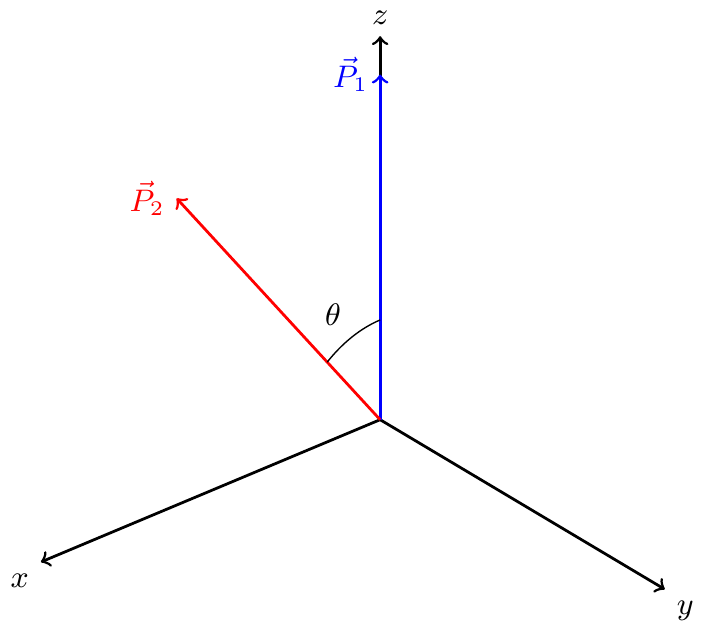}
\caption{Initial ($t = 0$) configuration of polarization vectors in the pure-self-coupling scenario.  For convenience the coordinate system is chosen such that $\vec{P}_1$ lies along the $z$-axis.  \label{purefig}}
\end{figure}

Both flavor amplitudes must individually satisfy Eq.~\eqref{pureosc}, which has solutions that are superpositions of $e^{\lambda_+ t}$ and $e^{\lambda_- t}$ with
\begin{equation}
\lambda_\pm = i \frac{\mu_1 - \mu_2}{2} \left[ 1 \pm \sqrt{1 + 4 \frac{\mu_1 \mu_2}{\left( \mu_1 - \mu_2 \right)^2} | \langle \psi_2 (0) | \psi_1 (0) \rangle |^2} \right].
\end{equation}
Note that the signs of the eigenvalues depend on whether $\mu_1$ or $\mu_2$ is larger.  In both cases we are letting $\lambda_+$ denote the eigenvalue of greater magnitude.  Thus
\begin{align}
& \psi_1 (t) = \left( \begin{array}{c}
a_+ e^{\lambda_+ t} + a_- e^{\lambda_- t} \\
b_+ e^{\lambda_+ t} + b_- e^{\lambda_- t}
\end{array} \right), \notag \\
& \psi_2 (t) = \left( \begin{array}{c}
c_+ e^{-\lambda_+ t} + c_- e^{-\lambda_- t} \\
d_+ e^{-\lambda_+ t} + d_- e^{-\lambda_- t}
\end{array} \right). \label{puregensolns}
\end{align}
With pure self-coupling, the mass axis is irrelevant and we are free to choose more convenient coordinates than those used in the previous section.  We let $\vec{P}_1 (0)$ point along the $z$-axis, and we let $\vec{P}_2 (0)$ be at an angle $\theta$ (Fig.~\ref{purefig}).  Then $| \langle \psi_1 (t) | \psi_2 (t) \rangle |^2 = \cos^2 \frac{\theta}{2}$ and the coefficients in Eq.~\eqref{puregensolns} are fixed by the parameters of the system.

%\begin{figure}
%\tdplotsetmaincoords{60}{140}
%\begin{tikzpicture}[scale=.9,tdplot_main_coords]
%
%\draw[thick,->] (0,0,0) -- (5,0,0) node[anchor=north east]{$x$};
%\draw[thick,->] (0,0,0) -- (0,5,0) node[anchor=north west]{$y$};
%\draw[thick,->] (0,0,0) -- (0,0,5) node[anchor=south]{$z$};
%
%\pgfmathsetmacro{\ax}{0}
%\pgfmathsetmacro{\ay}{0}
%\pgfmathsetmacro{\az}{4.5}
%
%\draw[thick,->,blue] (0,0,0) -- (\ax,\ay,\az) node[anchor=east]{$\vec{P}_1$};
%\draw[thick,->,red](0,0,0) -- (3,0,4) node[anchor=east]{$\vec{P}_2$};
%
%\tdplotgetpolarcoords{3}{0}{4}
%\tdplotsetthetaplanecoords{\tdplotresphi}
%\tdplotdrawarc[tdplot_rotated_coords]{(0,0,0)}{1.3}{0}%
%{\tdplotrestheta}{anchor=south east}{$\theta$}
%
%\end{tikzpicture}
%\caption{Initial ($t = 0$) configuration of polarization vectors in the pure-self-coupling scenario.  For convenience the coordinate system is chosen such that $\vec{P}_1$ lies along the $z$-axis. \label{purefig}}
%\end{figure}

We now pose this question: If $H_1$ undergoes cyclic evolution, does $| \psi_2 \rangle$ acquire a geometric phase?  Given the structure of the solutions in Eq.~\eqref{puregensolns} it is clear that $H_1$ and $H_2$ both cycle after a shared period $T$.  The same question can then be asked of $| \psi_1 \rangle$ with respect to cyclic evolution of $H_2$, and the geometric phases that emerge in this scenario must in some sense be coupled to one another.  Based on the solutions found above, $| \psi_1 \rangle$ and $| \psi_2 \rangle$ each complete a cycle after a time
\begin{equation}
T = \frac{2 \pi}{\left| \mu_1 - \mu_2 \right| \sqrt{ 1 + 4 \frac{\mu_1 \mu_2}{\left( \mu_1 - \mu_2 \right)^2} \cos^2 \frac{\theta}{2}}},
\end{equation}
at which point the wavefunctions have acquired the phases $e^{i \alpha_1}$ and $e^{i \alpha_2}$, respectively, with $\alpha_2 = - \alpha_1$ and
\begin{equation}
\alpha_1 = \textrm{sgn} \left( \mu_1 - \mu_2 \right) \left[ \frac{\pi}{\sqrt{ 1 + 4 \frac{\mu_1 \mu_2}{\left( \mu_1 - \mu_2 \right)^2} \cos^2 \frac{\theta}{2}}} + \pi \right].
\end{equation}
These are the exact phases acquired by the states after a time $T$.  Their geometric structure is manifest, and they are clearly coupled, as one is the negative of the other regardless of the choice of system parameters.  The result is also notable in that the dynamical phase makes no appearance: Since there are neither external parameters tuning the system nor even internal parameters associated with vacuum oscillations, the only timescale available is the intrinsic dynamical one set by the neutrino densities and the initial flavor states.

If $\mu_1 = \mu_2$, then no observable phase results at all.  With the neutrino densities equal, the first-derivative term in Eq.~\eqref{pureosc} drops out and the eigenvalues are related by a sign change.  The result is that only trivial overall phases can develop over the course of a cycle.  It is also straightforward to show that in the extreme limit $\mu_1 \gg \mu_2$, the geometric phase acquired by $| \psi_1 \rangle$ reduces to
\begin{equation}
\alpha_1 \longrightarrow - 2 \pi \frac{\mu_2}{\mu_1} \cos^2 \frac{\theta}{2},
\end{equation}
and still $\alpha_2 = - \alpha_1$.  Using Eq.~\eqref{upsilondef}, an analysis like the one in the previous section shows that the $| \psi_2 \rangle$ adiabaticity parameter is $\Upsilon \sim \mu_1 / \mu_2$.  This limit therefore describes adiabatic evolution of the relatively dilute population of neutrinos.

The same result can be obtained through the usual adiabatic treatment, where the eigensystem is solved for and the gauge potentials are calculated.  To demonstrate this, we now assume adiabaticity and compute the eigenvectors of $H_2 = \mu_1 \rho_1$.  The first eigenvector is simply $| \nu_\mu (t) \rangle = | \psi_1 (t) \rangle$, with eigenvalue $\lambda_\mu = \mu_1$.  The second eigenvector $| \nu_0 (t) \rangle$ has eigenvalue $\lambda_0 = 0$ and satisfies $\langle \nu_0 (t) | \psi_1 (t) \rangle = 0$.  These can be written out as
\begin{align}
& | \nu_\mu \rangle = \left( \begin{array}{c}
a_+ e^{\lambda_+ t} + a_- e^{\lambda_- t} \\
b_+ e^{\lambda_+ t} + b_- e^{\lambda_- t}
\end{array} \right), \notag \\
& | \nu_0 \rangle = \left( \begin{array}{c}
- b_+^* e^{\lambda_+^* t} - b_-^* e^{\lambda_-^* t} \\
a_+^* e^{\lambda_+^* t} + a_-^* e^{\lambda_-^* t}
\end{array} \right),
\end{align}
from which the gauge potentials---now written as scalars in order to facilitate the computation---may be evaluated:
\begin{align}
& A_\mu =  i \bigg\langle \nu_\mu \bigg| \frac{d}{dt} \bigg| \nu_\mu \bigg\rangle = \mu_2 \cos^2\frac{\theta}{2} \\
& A_0 = i \bigg\langle \nu_0 \bigg| \frac{d}{dt} \bigg| \nu_0 \bigg\rangle = - \mu_2 \cos^2\frac{\theta}{2},
\end{align}
The computation of the first of these is significantly aided by using $i \langle \nu_\mu | \frac{d}{dt} | \nu_\mu \rangle = \langle \psi_1 | H_1 | \psi_1 \rangle$, and the second can then be obtained easily by confirming that $\langle \nu_0 | \frac{d}{dt} | \nu_0 \rangle = \langle \nu_\mu | \frac{d}{dt} | \nu_\mu \rangle ^*$.  Thus, to first order,
\begin{align}
& i \int_0^T dt \bigg\langle \nu_\mu \bigg| \frac{d}{dt} \bigg| \nu_\mu \bigg\rangle = 2 \pi \frac{\mu_2}{\mu_1} \cos^2\frac{\theta}{2} \notag \\
& i \int_0^T dt \bigg\langle \nu_0 \bigg| \frac{d}{dt} \bigg| \nu_0 \bigg\rangle = - 2 \pi \frac{\mu_2}{\mu_1} \cos^2\frac{\theta}{2}.
\end{align}
These results yield a geometric phase consistent with the expansion of the exact phase in the $\mu_1 \gg \mu_2$ limit.  Here we have exhibited phases that, while not purely geometric, nonetheless arise in addition to the dynamical phase.

It is in fact not immediately apparent that this adiabatic treatment, where the geometric phase is calculated from the gauge potentials, even \textit{should} give the correct result. To see why, consider that the ``off-diagonal'' matrix elements are
\begin{equation}
\bigg\langle \nu_0 \bigg| \frac{d}{dt} \bigg| \nu_\mu \bigg\rangle = -i \mu_2 \cos\frac{\theta}{2} \sin\frac{\theta}{2} e^{i \left( \mu_2 - \mu_1 \right) t}
\end{equation}
and $\langle \nu_\mu | \frac{d}{dt} | \nu_0 \rangle = - \langle \nu_0 | \frac{d}{dt} | \nu_\mu \rangle ^*$, which is to say that they do not vanish any faster in the small-$\mu_2$ limit than the diagonal gauge potentials do.  Evidently, however, one gets the correct results if these terms are simply dropped.  The reason is that if the state is purely $| \nu_\mu \rangle$ or $| \nu_0 \rangle$ at $t = 0$, then the component along the other eigenstate grows slowly by virtue of being driven by $\mu_2$.  This small component in turn contributes to the phase evolution of the dominant component with another factor of $\mu_2$.  Hence it is appropriate after all to ignore the overlap with the small component.

What we have shown in this section is that even away from the adiabatic limit phases arise that depend on (1) the number densities of the two neutrino populations and (2) the constant angle between the polarization vectors, but \textit{not} explicitly on the time over which the system is evolved.  Furthermore, the geometric phases associated with the two states are necessarily related. In contrast to what was found in the previous section, the geometric phases here are living creatures: $| \psi_1 \rangle$ and $| \psi_2 \rangle$ mutually settle, simultaneously, on their paths in flavor space and on the attendant phases.

\section{The $\mu \ll \omega$ limit with two flavors \label{weakselfsec}}

In Sec.~\ref{2mixedsec} we showed that a neutrino acquires a geometric phase when it is strongly coupled to another neutrino undergoing vacuum oscillations; under these circumstances the Hamiltonian acts like an external, time-dependent ``flavor-magnetic'' field.  In Sec.~\ref{puresec} we showed that geometric phases can survive when the evolution of the magnetic field is coupled back to the test neutrino.  We now ask whether geometric phases persist when vacuum oscillations and self-coupling are accounted for in both population of neutrinos.  In particular we consider geometric effects arising in the $\mu \ll \omega$ limit.

We return to Eq.~\eqref{starteom} and assume that the neutrino--neutrino forward-scattering potentials are small compared to the vacuum potentials.  To prepare to use perturbation theory, we write the equations of motion as
\begin{align}
& i \frac{d | \psi_1 \rangle}{dt} = \left[ \omega_1 B + \epsilon \mu_2 \rho_2 \right] | \psi_1 \rangle, \notag \\
& i \frac{d | \psi_2 \rangle}{dt} = \left[ \omega_2 B + \epsilon \mu_1 \rho_1 \right] | \psi_2 \rangle.
\end{align}
We expand perturbatively in the small parameter $\epsilon$:
\begin{align} 
& | \psi_1 \rangle = | \psi_1^{(0)} \rangle + \epsilon | \psi_1^{(1)} \rangle + \dots , \notag \\
& | \psi_2 \rangle = | \psi_2^{(0)} \rangle + \epsilon | \psi_2^{(1)} \rangle + \dots .
\end{align}
To zeroth order the equations of motion are just those for vacuum oscillation:
\begin{align}
& i \frac{d | \psi_1^{(0)} \rangle}{dt} = \omega_1 B | \psi_1^{(0)} \rangle, \notag \\
& i \frac{d | \psi_2^{(0)} \rangle}{dt} = \omega_2 B | \psi_2^{(0)} \rangle,
\end{align}
which have solutions
\begin{align}
& | \psi_1^{(0)}(t) \rangle = \exp \left( -i \omega_1 B t \right) | \psi_1^{(0)}(0) \rangle, \notag \\
& | \psi_2^{(0)}(t) \rangle = \exp \left( -i \omega_2 B t \right) | \psi_2^{(0)}(0) \rangle.
\end{align}
The first-order equation for $| \psi_1 \rangle$ is
\begin{equation}
i \frac{d | \psi_1^{(1)} \rangle}{dt} = \omega_1 B \psi_1^{(1)} + \mu_2 \rho_2^{(0)} | \psi_1^{(0)} \rangle,
\end{equation}
which, after plugging in the zeroth-order solution for $| \psi_2 \rangle$, becomes
\begin{align}
\frac{d | \psi_1^{(1)} \rangle}{dt} = & - i \omega_1 B | \psi_1^{(1)}(t) \rangle \notag \\
& - i \mu_2 e^{-i \omega_2 B t} \rho_2^{(0)}(0) e^{i \left( \omega_2 - \omega_1 \right) B t} | \psi_1^{(0)}(0) \rangle.
\end{align}
This has solution
\begin{align}
| \psi_1^{(1)}(t) \rangle &= e^{-i \omega_1 B t} \bigg[ | \psi_1^{(1)}(0) \rangle - i \mu_2 \notag \\
& \times \int_0^t dt' e^{i \omega_1 B t'} \rho_2^{(0)}(t') | \psi_1^{(0)}(t') \rangle \bigg], \label{psi1soln}
\end{align}
and $| \psi_2^{(1)}(t) \rangle$ has an identical form but with subscripts interchanged.

Cyclicity fails to materialize as naturally here as it did in earlier sections.  To find geometric effects analogous to the ones reported above, we seek values of the period $T$ such that
\begin{equation}
| \psi_2(T) \rangle = e^{i\alpha} | \psi_2(0) \rangle. \label{psi2alpha}
\end{equation}
The phase $\alpha$ is to be solved for concomitantly.  Since $| \psi_1 \rangle$ may not be cyclic with the same period, in general $\alpha$ will not be a phase of the Berry genus.

The phase and period are expanded as
\begin{align}
& \alpha = \alpha^{(0)} + \epsilon \alpha^{(1)}, \notag \\
& T = T^{(0)} + \epsilon T^{(1)}.
\end{align}
If $\vec{P}_2$ is initially at angles $( \theta_2, \phi_2 )$, then the initial conditions for this mode are
\begin{align}
& | \psi_2^{(0)}(0) \rangle = \left( \begin{array}{c}
\cos\frac{\theta_2}{2} \\
e^{i \phi_2} \sin\frac{\theta_2}{2}
\end{array} \right), \notag \\
& | \psi_2^{(1)}(0) \rangle = \left( \begin{array}{c}
0 \\
0
\end{array} \right).
\end{align} 
Demanding that $| \psi_2 \rangle$ satisfy Eq.~\eqref{psi2alpha} then amounts to the following requirements on $\alpha$ and $T$:
\begin{align}
& | \psi_2^{(0)} (T^{(0)}) \rangle = e^{i \alpha^{(0)}} | \psi_2 ^{(0)} (0) \rangle, \notag \\
& | \psi_2^{(1)} (T^{(0)}) \rangle + T^{(1)} \frac{ d | \psi_2 ^{(0)} \rangle}{dt} \bigg|_{T^{(0)}} = i \alpha^{(1)} e^{i \alpha^{(0)}} | \psi_2 ^{(0)} (0) \rangle.
\end{align}
The first equation is satisfied if
\begin{equation}
T^{(0)} = \frac{2 \pi n}{\omega_2}, ~~~ \alpha^{(0)} = n \pi,
\end{equation}
with $n \in \mathbb{Z}$.  The second equation then becomes
\begin{equation}
\bigg| \psi_2^{(1)} \left( \frac{2 \pi n}{\omega_2} \right) \bigg\rangle = \pm i \left( \alpha^{(1)} + \omega_2 T^{(1)} B \right) | \psi_2 ^{(0)} (0) \rangle,
\end{equation}
where $| \psi_2^{(1)} \rangle$ can be evaluated using Eq.~\eqref{psi1soln}.  The $+$ ($-$) corresponds to even (odd) $n$.  

In deriving the first-order corrections to the phase and period it is helpful to note a few intermediate results.  First, the term in the integrand of Eq.~\eqref{psi1soln} that multiplies $| \psi_1^{(0)}(t) \rangle$ is
\begin{widetext}
\begin{equation}
e^{i \left( \omega_2 - \omega_1 \right) B t} \rho_1^{(0)}(0) e^{- i \left( \omega_2 - \omega_1 \right) B t} = \left( \begin{array}{cc}
\cos^2 \frac{\theta_1}{2} & e^{i \left( \omega_1 - \omega_2 \right) t} e^{-i \phi_1} \cos\frac{\theta_1}{2} \sin\frac{\theta_1}{2} \\
e^{-i \left( \omega_1 - \omega_2 \right) t} e^{i \phi_1} \cos\frac{\theta_1}{2} \sin\frac{\theta_1}{2} & \sin^2\frac{\theta_2}{2}
\end{array} \right).
\end{equation}
\end{widetext}
The two conditions that can be extracted from the matrix solution for $| \psi_1^{(1)} (t) \rangle$ are then
\begin{align}
& 2 \pi n \frac{\mu_1}{\omega_2} \cos^2\frac{\theta_1}{2} + \xi ^* \tan\frac{\theta_2}{2} = - \alpha^{(1)} + \frac{\omega_2 T^{(1)}}{2}, \notag \\
& 2 \pi n \frac{\mu_1}{\omega_2} \sin^2\frac{\theta_1}{2} + \xi \cot\frac{\theta_2}{2} = - \alpha^{(1)} - \frac{\omega_2 T^{(1)}}{2},
\end{align}
where
\begin{equation}
\xi \equiv i \mu_1 \frac{e^{-2 \pi n i \left( \frac{\omega_1}{\omega_2} - 1 \right)} - 1}{\omega_1 - \omega_2} e^{i \left(\phi_1 - \phi_2 \right)} \cos\frac{\theta_1}{2} \sin\frac{\theta_1}{2}.
\end{equation}
These equations can be solved to yield
\begin{align}
& \alpha^{(1)} = - n \pi \frac{\mu_1}{\omega_2} - \frac{\xi}{2} \cot \frac{\theta_2}{2} - \frac{\xi ^*}{2} \tan\frac{\theta_2}{2}, \notag \\
& T^{(1)} = \frac{2 \pi n}{\omega_2}\frac{\mu_1}{\omega_2} \cos\theta_1 + \frac{\xi ^*}{\omega_2} \tan \frac{\theta_2}{2} - \frac{\xi}{\omega_2} \cot\frac{\theta_2}{2}.
\end{align}
Notice that in general $\xi$ is complex.  Demanding that the period be real (but without putting contrived restrictions on the angles) requires that $\omega_1 / \omega_2$ be a rational number of the form $m / n$, with $m \in \mathbb{Z}$.  Choosing $\omega_1$ to satisfy this constraint, one obtains
\begin{align}
& \alpha^{(1)} = - \pi n \frac{\mu_1}{\omega_2}, \notag \\
& T^{(1)} = 2 \pi n \frac{\mu_1}{\omega_2^2} \cos\theta_1,
\end{align}
so that to first order we have
\begin{align}
& T = \frac{2 \pi n}{\omega_2} \left( 1 + \frac{\mu_1}{\omega_1} \cos\theta_1 \right), \notag \\
& \alpha = n \pi \left( 1 - \frac{\mu_1}{\omega_2} \right).
\end{align}
To ensure that $| \psi_2 \rangle$ is cyclic, the rationality condition $\omega_1 / \omega_2 = m / n$ is necessary---but having so picked the vacuum oscillation frequencies, $| \psi_2 \rangle$ oscillates with a geometry-dependent period and acquires a phase sensitive to the density of the other neutrino population.

Analogous results apply if instead we take $| \psi_1 \rangle$ to be cyclic and seek out the period and phase consistent with such a requirement.  If $\mu_1 = \mu_2$ and $\omega_1 = \omega_2$, then $| \psi_1 \rangle$ and $| \psi_2 \rangle$ are cyclic with the same period and accrue identical phases.  In this particular scenario, where the two neutrino populations consist of particles of the same energy and density, the phases are of the classic Berry type, with each population experiencing adiabatic evolution under a cyclic Hamiltonian.  In general the adiabaticity parameter for $| \psi_1 \rangle$ is $\Upsilon_1 \sim \omega_1^2 / \mu_2 | \omega_2 |$, and similarly for $| \psi_2 \rangle$.  Adiabaticity is therefore established automatically by taking the limit $\mu \ll \omega$, so long as the frequencies are of comparable magnitude.  This observation also matches intuition: The time-dependent self-coupling potential, which elicits deviations from adiabaticity, is only a small part of the total Hamiltonian.

%For unequal densities and frequencies, $| \psi_1 \rangle$ and $| \psi_2 \rangle$ have different periods.  We now ask what happens to $| \psi_1 \rangle$ after a time $T$ such that $| \psi_2 \rangle$ is cyclic.  To be consistent with the order of expansion, we use
%\begin{equation}
%\psi_1 (T) = \psi_1^{(0)}\left( T^{(0)} + T^{(1)} \right) + \psi_1^{(1)}\left( T^{(0)} \right).
%\end{equation}
%One finds
%\begin{equation}
%| \psi_1 (T) \rangle = \left( \begin{array}{c}
%\mathcal{A} \cos\frac{\theta_1}{2} \\
%\mathcal{B} e^{i \phi_1} \sin \frac{\theta_1}{2}
%\end{array} \right),
%\end{equation}
%where the coefficients are
%\begin{align}
%& \mathcal{A} = e^{n \pi i \frac{\omega_1}{\omega_2} } \left( e^{n \pi i \frac{\mu_1}{\omega_2} \cos \theta_1} - 2 \pi n i \frac{\mu_2}{\omega_2} \right) \notag \\
%& \mathcal{B} = e^{- n \pi i \frac{\omega_1}{\omega_2} } \left( e^{- n \pi i \frac{\mu_1}{\omega_2} \cos \theta_1} - 2 \pi n i \frac{\mu_2}{\omega_2} \right)
%\end{align}
%Given the rationality condition on the frequencies, the prefactors outside the parentheses contribute a $\pm$ to both $\mathcal{A}$ and $\mathcal{B}$.  Asking that $\psi_1$ be cyclic to this order as well then means that
%\begin{equation}
%\frac{\mu_1}{\omega_2} \cos\theta_1 = \frac{l}{n},
%\end{equation}
%where $l$ is any integer.  This is yet another contrived condition.

In the final assessment, cyclicity is typically jeopardized when the nonlinear coupling acts to perturb the neutrinos away from vacuum oscillations.    Nonetheless geometry remains relevant to the flavor transformation that occurs in such a system, as evidenced by the noncyclic variants of the geometric phase already alluded to in Sec.~\ref{introsec} (see, \textit{e.g.}, Refs.~\cite{wang2001, mehta2009} for applications to neutrinos in vacuum+matter).  We do not pursue this direction any further but we do emphasize that the imprints of geometry in flavor transformation transcend the cyclic, adiabatic phase.

\section{Mixed potentials with three flavors: non-Abelian phase \label{3mixedsec}}

We now generalize the mixed-potentials scenario of Sec.~\ref{2mixedsec} to three flavors.  As before, the vacuum oscillations of one neutrino determine the Hamiltonian experienced by the other.  That is,
\begin{align}
& i \frac{d | \psi_1 (t) \rangle}{dt} = \mu_2 \rho_2 (t) | \psi_1 (t) \rangle, \notag \\
& i \frac{d | \psi_2 (t) \rangle}{dt} = H_{\textrm{vac}, 2}  | \psi_2 (t) \rangle,
\end{align}
where $| \psi_i \rangle$ is a three-component vector, $\rho_i$ is a $3 \times 3$ matrix, and in the mass basis
\begin{equation}
H_{\textrm{vac}, 2} = \frac{1}{3} \left( \begin{array}{ccc}
-\Delta_{21} - \Delta_{31} & 0 & 0 \\
0 & \Delta_{21} - \Delta_{32} & 0 \\
0 & 0 & \Delta_{32} + \Delta_{31}
\end{array} \right),
\end{equation}
using the notation $\Delta_{ij} \equiv \delta m_{ij}^2 / 2 E$.  The equations of motion have solutions
\begin{align}
& | \psi_1 (t) \rangle = \mathcal{P} \exp \left( -i \mu_1 \int_0 ^t dt' \rho_2 (t') \right) | \psi_1 (0) \rangle, \notag \\
& | \psi_2 (t) \rangle = \exp \left( -i H_{\textrm{vac}, 2} t \right) | \psi_2 (0) \rangle.
\end{align}
The matrix exponential for the second of these is straightforward to compute.  The solution, for $| \psi_2 (0) \rangle = ( a, b, c )^T$, is
\begin{equation}
| \psi_2 (t) \rangle = \left( \begin{array}{c}
\exp \left( i \frac{\Delta_{21}+\Delta_{31}}{3} t \right) a \\
\exp \left( i \frac{\Delta_{32}-\Delta_{21}}{3} t \right) b \\
\exp \left( -i \frac{\Delta_{32}+\Delta_{31}}{3} t \right) c
\end{array} \right).
\end{equation}
It follows that
%\begin{widetext}
\begin{align}
& \rho_2 (t) = \notag \\
& \left( \begin{array}{ccc}
| a |^2 & \exp \left( i \Delta_{21} t \right) ab^* & \exp \left( i \Delta_{31} t \right) ac^* \\
\exp \left( - i \Delta_{21} t \right) ba^* & | b |^2 & \exp \left( i \Delta_{32} t \right) bc^* \\
\exp \left( -i \Delta_{31} t \right) ca^* & \exp \left( - i \Delta_{32} t \right) cb^* & | c |^2
\end{array} \right)
\end{align}
%\end{widetext}
and thus the geometric phases induced by the Hamiltonian $H_1 (t) = \mu_2 \rho_2 (t)$ can be found by solving for the eigensystem.  (In carrying out this procedure, one is aided by the Cardano formula.)  The eigenvalues are $E_\mu = \mu_2$, $E_0 = 0$, and $E_{0'} = 0$, which correspond respectively to the eigenvectors
\begin{align}
& | \nu_\mu \rangle = \frac{1}{|c|} \left( \begin{array}{c}
\exp \left( i \Delta_{31} t \right) ac^* \\
\exp \left( i \Delta_{32} t \right) bc^* \\
|c|^2
\end{array} \right), \notag \\
& | \nu_0 \rangle = \frac{1}{\sqrt{1 + \frac{|c|^2}{|b|^2}}} \left( \begin{array}{c}
\exp \left( i \Delta_{31} t \right) ac^* \left( 1 - \frac{1}{|a|^2} \right) \\
\exp \left( i \Delta_{32} t \right) bc^* \\
|c|^2
\end{array} \right), \notag \\
& | \nu_{0'} \rangle = \frac{|a|^2}{|c|^2 \sqrt{1-|a|^2}} \left( \begin{array}{c}
0 \\
-\exp \left( i \Delta_{32} t \right) \frac{c^*}{b^*} \\
1
\end{array} \right).
\end{align}
In this scenario two of the energy eigenstates are always degenerate, indicating that the geometric phases have a non-Abelian gauge structure \cite{wilczek1984}.

In the two-flavor case, where the gauge was Abelian, the geometric phases acquired by the energy eigenstates could be deduced by solving the Schr\"odinger equation with $| \psi_\pm (t) \rangle = \exp \left( i \phi_\pm (t) \right) | \nu_\pm (t) \rangle$; this is the condition that enforces perfect adiabaticity.  But nothing prevents the states within the degenerate subsystem from mixing with each other, regardless of how adiabatic the evolution is.  To find the phases in the degenerate subsystem, one must therefore solve the Schr\"odinger equation with $| \psi _i (t) \rangle = U _{ij} (t) | \nu _j (t) \rangle$, where $| \nu _j (t) \rangle$ is the $j^\textrm{th}$ eigenstate and $U(t)$ is the matrix generalizing the Abelian phase from the two-flavor case.  If the path over a time $t$ corresponds to a closed loop $\mathcal{C}$, then the matrix is given by the Wilson loop
\begin{equation}
U (\mathcal{C}) = \mathcal{P} \exp \left( i \oint _\mathcal{C} \vec{A} \cdot d\vec{R} \right), \label{nonabelianu}
\end{equation}
where the gauge potential is now a vector-valued matrix with components
\begin{equation}
\vec{A}_{ij} = i \big\langle \nu _i (t) \big| \nabla \big| \nu _j (t) \big\rangle .
\end{equation}
Eq.~\eqref{nonabelianu} generalizes Eq.~\eqref{gammadef}.  We do not write out all of the gauge potentials since they are not particularly enlightening, but we do note that the off-diagonal elements of $A$ are nonzero, allowing for transitions between $| \nu_0 \rangle$ and $| \nu_{0'} \rangle$ even in the adiabatic limit.  If $| \psi_2 (0) \rangle = | \nu_\alpha \rangle$ for $\alpha = e, \mu, \tau$, the transitions occur between orthogonal linear combinations of the other two flavors.

For $H_1$ to be cyclic, the period must be an integer multiple, all at once, of $2 \pi / \Delta_{21}$, $2 \pi / \Delta_{31}$, and $2 \pi / \Delta_{32}$.  This reflects the requirement for a three-flavor neutrino to oscillate back into its original state in finite time, a condition that was guaranteed in the two-flavor case.  Supposing that such a $T$ does exist, one can show that indeed the phases arising from the gauge potentials do not depend explicitly on time.

The non-Abelian structure owes its existence to a basic fact about the Hamiltonian for $| \psi_1 \rangle$.  It is a fact that appeared previously in our study of the pure-self-coupling scenario in two flavors: When $H \sim \rho = | \psi \rangle \langle \psi |$, one eigenstate is $| \psi \rangle$ itself and all others are orthogonal states with eigenvalue 0.  (Note that for the mixed-potential scenario in two flavors, where this trait of the Hamiltonian was also relevant, we pulled out the trace of the self-coupling Hamiltonian and thereby shifted the orthogonal-eigenstate energy down to $-\mu_1 / 2$.)  To put it more starkly, the non-Abelian phase structure is a consequence of the coupling of neutrino flavor quantum states---with off-diagonal coherence included---rather than merely neutrino flavor number densities.

%These considerations hold regardless of the number of flavors, so that with $N$ active flavors one would have a projective Hilbert space $SU(N) / U(N-1)$.  \textbf{[of course, ruled out]}

%It is expected that the inclusion of a vacuum term will reduce this symmetry to an Abelian one, unless perhaps there is a mass degeneracy.

\section{Conclusion \label{concsec}}

We have pointed out that the self-coupling potential generated by neutrino--neutrino forward scattering is capable of inducing geometric phases in flavor evolution.  The mechanism is most easily understood in the two-flavor approximation, where a neutrino's flavor state and Hamiltonian correspond graphically to vectors ending on the Bloch sphere.  In a background consisting strictly of matter particles, the Hamiltonian vector $\vec{H}$ is confined to a plane.  But in a medium dense in neutrinos, $\vec{H}$ is liberated from the plane and, should it undergo a closed cycle, may return to its initial point having enclosed a finite solid angle on the sphere.  Path-dependent geometric phases in the energy eigenstates are the result---and since flavor transformation at its heart is an interference phenomenon of the neutrino's energy eigenstates, the phases surface in flavor transition probabilities and are observable in principle.

To examine these phases in an analytically tractable setting, we have considered various limits of a very simple toy model devoid of the astrophysical complications that beckon a numerical treatment.  Despite the model's simplicity, the calculations presented in this paper illuminate several facets of geometric phases in environments with nonlinear refraction from neutrino self-coupling.

Foremost among these aspects are the roles of adiabaticity and cyclicity.  We have seen that adiabatic evolution is not a necessity, and that geometric effects are apparent in the non-adiabatic corrections, albeit in a way entangled with the dynamics.  We have also seen that the complicated interplay between oscillations and self-coupling tends to compromise cyclicity.  But cyclicity is also dispensable, and though we have not pursued this direction here, it is expected that geometric effects should prove to be a generic feature of noncyclic evolution as well.

Beyond these, two other interesting phenomena have emerged from the calculations: the entwining of the paths and phases of the two neutrino populations, as exhibited in the pure-self-coupling scenario, and the non-Abelian phase structure of the three-flavor case.  These effects hinge on the peculiar nature of the neutrino--neutrino forward-scattering potential, which allows neutrinos to communicate to one another the quantum coherence of their flavor states.

This study was motivated by the possibility for collective flavor-transformation effects in the extreme environments found, for instance, in the torrid plasma of the early universe or the incendiary outflow from a core-collapse supernova.  We have made no attempt to locate geometric phases in astrophysically realistic models but have instead strived to make clear, based on calculations in uncluttered toy models, how such phases might emerge.  Indeed, we expect that the ideas underlying this study may find a place, in some form, in a variety of applications: in synchronized or bipolar oscillations in the early universe, in a possibly cyclic halo-affected region outside a supernova, in active--sterile oscillations, and elsewhere.  To be sure, sophisticated numerical computations already have geometric effects built in implicitly, albeit in far more complicated manifestations than those analyzed here.  After all, the provenance of these effects---the shape of Hilbert space and the structure of the Hamiltonian---is encoded in the equations of motion.  But the importance of geometry in the results that these equations output is often overlooked.

\begin{acknowledgments}
We thank Amol Patwardhan for helpful conversations.  This work was supported by NSF Grant No. PHY-1307372 at UC San Diego.  The majority of the project was completed while one of us (L.J.) was stationed at Los Alamos National Laboratory, with support from the U.S. Department of Energy Office of Science Graduate Student Research (SCGSR) Program.  This author thanks the Lab for their hospitality.
\end{acknowledgments}

%\bibliographystyle{abbrv}
\bibliography{all_papers}

\end{document}